\newcommand*{\ATLASLATEXPATH}{latex/}
\documentclass[cernpreprint,texlive=2011,txfonts,UKenglish]{\ATLASLATEXPATH atlasdoc}
\pdfoutput=1

\usepackage{\ATLASLATEXPATH atlaspackage}
\usepackage{\ATLASLATEXPATH atlasbiblatex}

\usepackage{\ATLASLATEXPATH atlascontribute}

\usepackage[hion]{\ATLASLATEXPATH atlasphysics}

\graphicspath{{logos/}{figures/}}

\usepackage{MB8-defs}

\AtlasTitle{Charged-particle distributions in $pp$ interactions at $\sqrt{s}=8 \TeV$ measured with the ATLAS detector}

\author{The ATLAS Collaboration}

\AtlasContributor{Wayne~Cribbs}{contribution}
\AtlasContributor{Oldrich~Kepka}{contribution}
\AtlasContributor{Thorsten~Kuhl}{contribution}
\AtlasContributor{Wolfgang~Lukas}{contribution}
\AtlasContributor{Roman~Lysak}{contribution}
\AtlasContributor{Morten~Ankersen~Medici}{contribution}
\AtlasContributor{David~Milstead}{contribution}
\AtlasContributor{Simone~Pagan~Griso}{contribution}
\AtlasContributor{Andreas~Salzburger}{contribution}
\AtlasContributor{Edward~Sarkisyan-Grinbaum}{contribution}
\AtlasContributor{Krzysztof~Wozniak}{contribution}

\AtlasRefCode{STDM-2014-19}

\PreprintIdNumber{CERN-EP-2016-020}

\arXivId{1603.02439}

\HepDataRecord{ins1426695}

\AtlasJournal{\EPJC}
\AtlasJournalRef{\EPJC{} 76 (2016) 403}
\AtlasDOI{10.1140/epjc/s10052-016-4203-9}

\AtlasAbstract{%
  This paper presents measurements of distributions of charged particles which are produced in proton--proton collisions at a centre-of-mass energy of $\sqrt{s} = 8 \TeV$ and 
recorded by the ATLAS detector at the LHC.
  A special dataset recorded in 2012 with a small number of interactions per beam crossing (below 0.004) and corresponding to an integrated luminosity of \SI{160}{\per\micro\barn} was used.
  A minimum-bias trigger was utilised to select a data sample of more than 9 million collision events.
  The multiplicity, pseudorapidity, and transverse momentum distributions of charged particles are shown in different regions of kinematics and charged-particle multiplicity, including measurements of final states at high multiplicity.
  The results are corrected for detector effects and are compared to the predictions of various Monte Carlo event generator models which 
simulate the full hadronic final state.
}

\AtlasCoverSupportingNote{Charged-particle multiplicities in $pp$ interactions at $\sqrt{s}=8 \TeV$ measured with the ATLAS detector at the LHC}{https://cds.cern.ch/record/1700840}

\AtlasCoverCommentsDeadline{31 January 2016}

\AtlasCoverAnalysisTeam{Wayne~Cribbs, Oldrich~Kepka, Thorsten~Kuhl, Wolfgang~Lukas, Roman~Lysak, Morten~Ankersen~Medici, David~Milstead, Simone~Pagan~Griso, Andreas~Salzburger, Edward~Sarkisyan-Grinbaum, Krzysztof~Wozniak}

\AtlasCoverEdBoardMember{Gerhard~Brandt~(chair)}
\AtlasCoverEdBoardMember{Konstantinos~Bachas}
\AtlasCoverEdBoardMember{Andy~Buckley}
\AtlasCoverEdBoardMember{Heiko~Lacker}
\AtlasCoverEdBoardMember{Tim~Martin}

\AtlasCoverEgroupEditors{atlas-stdm-2014-19-editors@cern.ch}

\AtlasCoverEgroupEdBoard{atlas-stdm-2014-19-editorial-board@cern.ch}

\hypersetup{pdftitle={Charged-particle distributions in pp interactions at sqrt(s) = 8 TeV measured with the ATLAS detector},pdfauthor={The ATLAS Collaboration}}

\begin{document}

\maketitle



\section{Introduction}
\label{sec:intro}

Measurements of charged-particle spectra probe strong interactions at low momentum transfers.
Such measurements have been made
in lower-energy $e^+e^-$, $ep$ and hadron collisions~\cite{Arnison:1982ed,Breakstone:1983ns,Ansorge:1986xq,Alner:1987wb,Ansorge:1988fg,Abe:1988yu,Ansorge:1988kn,Albajar:1989an,Abe:1989td,Alexopoulos:1994ag,Aaltonen:2009ne}
and at the CERN Large Hadron Collider (LHC)~\cite{Aamodt:2009aa,Khachatryan:2010xs,Aad:2010rd,Aamodt:2010pp,Khachatryan:2010us,Khachatryan:2010nk,Aad:2010ac,Aaij:2014pza,Chatrchyan:2014qka,Khachatryan:2015jna,Adam:2015pza,Aad:2016mok}.
This paper presents measurements of multiplicity distributions, as well as transverse momentum and pseudorapidity spectra, for primary charged particles produced in $pp$ collisions recorded by the ATLAS experiment~\cite{PERF-2007-01} at the LHC
at $8 \TeV$ centre-of-mass energy.

Although a description of low-energy processes within a perturbative framework is not possible, predictions can be made with phenomenological models inspired by quantum chromodynamics (QCD). Data are used to constrain such models and gain further insight into the particle dynamics of the low transverse momentum regime. Furthermore, low-\pT processes, arising from pile-up in which there is more than one interaction per beam crossing, may also affect the topologies of events involving an interaction with a high \pT scale. An understanding of soft QCD processes is therefore important both in its own right and as a means of reducing systematic uncertainties in measurements of high-\pT phenomena.

The measurements presented in this paper use a methodology similar to that used at lower centre-of-mass energies at ATLAS~\cite{Aad:2010ac}. Events were selected from data taken in 2012 with a trigger overlapping with the acceptance of the tracking volume. This corresponds to a minimum-bias dataset based on inelastic $pp$ interactions. The term \emph{minimum bias} is taken to refer to trigger and event selections which are as unrestrictive as possible for the $pp$-induced final states. The integrated luminosity of the data sample under study is
\SI{160}{\per\micro\barn}.
Owing to improvements in understanding the material inside and around the ATLAS inner detector (ID),
the uncertainties in the measured spectra are reduced by as much as
\SIrange[range-phrase = --]{30}{50}{\percent}
compared to the analogous measurements at
$7 \TeV$ centre-of-mass energy~\cite{Aad:2010ac}.

The following distributions are measured:

	$$
	\dNdeta \ , \ \ \
	\dNdpT \ , \ \ \
	\dnchnorm \ , \ \ \
	\mathrm{and} \ \ \
	\meanpT \ \mathrm{vs} \ \nch \ .
	$$

Here, $\eta$ is the particle's pseudorapidity,\footnote{%
\AtlasCoordFootnote} %
\pT is the component of the charged-particle momentum which is transverse to the beam direction,\footnote{%
The factor $2 \pi \pT$ in the \pT spectrum comes from the Lorentz-invariant definition of the cross-section in terms of ${\mathrm{d}^3}\hspace{-0.01cm}p$.
The results could thus be interpreted as the massless approximation to ${\mathrm{d}^3}\hspace{-0.01cm}p$.} %
\nch is the number of primary charged particles in an event,
\Nev is the event yield for a given event selection, 
and \Nch is the total number of primary charged particles in all selected events in the data sample.
A primary charged particle is defined as a charged particle with a mean lifetime
$\tau > \SI{300}{ps}$,
which is either
directly produced in $pp$ interactions or from decays of directly produced particles with $\tau < \SI{30}{\ps}$; particles
produced from decays of particles with $\tau > \SI{30}{\ps}$ are considered as secondary particles and are thus excluded.
Primary charged particles are furthermore required to satisfy the kinematic selection criteria of
$|\eta| < 2.5$ and either $\pT > 100 \MeV$ or $500 \MeV$.

In order to make a more complete study of particle properties in minimum-bias events, results are given for different multiplicity and kinematic selections (termed \emph{phase spaces}).
The most inclusive phase spaces correspond to events with a minimum multiplicity $\nch \geq 2$ or 1 and contain primary charged particles possessing a minimum transverse momentum $\pT > 100 \MeV$ or $500 \MeV$, respectively.
Primary-charged-particle spectra are also shown for higher-multiplicity events ($\pt > 500 \MeV$, $\nch \geq 6,20$ and $50$) of which the latter two event types have hitherto not been measured by ATLAS.
Finally, the average primary-charged-particle densities at central pseudorapidity are compared to existing measurements at different centre-of-mass energies.

\section{ATLAS detector}
\label{sec:detector}

The ATLAS detector
covers almost the whole solid angle around the collision point with layers of tracking detectors, calorimeters and muon chambers.
The tracking modules and the trigger system are of most relevance for the presented measurements.

The inner detector has full coverage in $\phi$ and covers the pseudorapidity range $|\eta| < 2.5$.
It comprises a silicon pixel detector (Pixel), a silicon microstrip detector (SCT) and a transition radiation tracker (TRT).
These detectors cover a sensitive radial distance from the interaction point of
\SIrange[range-phrase = --]{50.5}{150}{\mm},
\SIrange[range-phrase = --]{299}{560}{\mm} and
\SIrange[range-phrase = --, group-digits = false]{563}{1066}{\mm}, respectively, and are immersed in a \SI{2}{\tesla} axial magnetic field provided by a solenoid.
The inner-detector barrel (end-cap) parts consist of 3 ($2\times3$) Pixel layers, 4 ($2 \times 9$) double-layers of single-sided silicon microstrips with a \SI{40}{\mrad} stereo angle, and 73 ($2 \times 160$) layers of TRT straws.
Typical position resolutions are
\SIlist{10;17;130}{\um}
for the $r$--$\phi$ co-ordinate and, in the case of the Pixel and SCT,
\SIlist{115;580}{\um}
for the second measured co-ordinate.
A track from a primary charged particle traversing the barrel detector would typically have 11 silicon hits~\footnote{%
A hit is a measurement point in a sensitive detector element which can be assigned to a reconstructed track.} %
(3 pixel clusters and 8 strip clusters) and more than 30 TRT straw hits.

The ATLAS detector has a three-level trigger system: Level~1~(L1), Level~2~(L2) and Event~Filter~(EF).
For the presented measurements, the trigger relies on the L1 signals from the minimum-bias trigger scintillators (MBTS).
The MBTS are positioned at each end of the detector in front of the liquid-argon end-cap calorimeter cryostats at $z = \pm \SI{3.56}{m}$.
They are segmented into eight sectors in azimuth and two rings in pseudorapidity and cover the range $2.08 < |\eta| < 3.75$.
The MBTS triggers are configured to require at least one or two hits above threshold from either side of the detector.\footnote{%
In order to keep the readout rates from the MBTS trigger streams approximately constant during the run in which the data were collected, pre-scale factors that evolved during the run were applied for each trigger. Therefore, a combination of both L1 MBTS triggers was used to maximise the data yield and reduce statistical uncertainties.} %

\section{Monte Carlo simulation}
\label{sec:mc}

The following Monte Carlo~(MC) models of inclusive hadron--hadron interactions were used to generate event samples.
These models employ different settings of model parameters (referred to as \emph{tunes}) which were optimised to reproduce existing experimental data.

\begin{itemize}

   \item \PYTHIA 8~\cite{Sjostrand:2007gs} and \PYTHIA 6~\cite{Sjostrand:2006za}.
   In these models, the total inelastic cross-section is separated into non-diffractive~(ND) processes, dominated by $t$-channel gluon exchange, and diffractive processes where a colour-singlet object is exchanged.
   Multiple parton--parton interactions (MPI) contribute to multiplicity fluctuations and are simulated as part of the ND processes.
   The diffractive processes consist of single-diffractive dissociation~(SD) and double-diffractive dissociation~(DD).
   \PYTHIA 8 is used with the \textsc{A2}~\cite{ATLAS:2011krm} and \textsc{Monash}~\cite{Skands:2014pea} tunes. The \textsc{A2} tune was performed on minimum-bias and underlying-event data, utilising the \textsc{MSTW2008 LO}~\cite{Martin:2009iq} parton distribution function (PDF). The \textsc{Monash} tune was made using a re-analysis of fragmentation-sensitive measurements with $e^+e^-$ collisions, combined with minimum-bias and underlying-event tuning for hadron--hadron data, utilising the \textsc{NNPDF23LO} PDF.
   \PYTHIA 6 employs the \textsc{AMBT2B}~\cite{ATLAS:2011zja} tune with the \textsc{CTEQ6L1}~\cite{Pumplin:2002vw} PDF.
   The \textsc{AMBT2B} tune was evaluated using jet and minimum-bias data.

   \item \EPOS~\cite{Porteboeuf:2010um}.
   This model implements a parton-based Gribov--Regge theory~\cite{Drescher:2000ha}, which is an effective QCD-inspired field theory describing hard and soft scattering simultaneously.
   \EPOS has been primarily designed for \PbPb interactions and cosmic air showers.
   The \textsc{LHC} tune~\cite{Pierog:2013ria} is used here, which modifies the modelling of radial flow to be more applicable for small-volume, high-density regions, as are found in $pp$ interactions.

   \item \QGSJET~\cite{Ostapchenko:2010vb} using the default tune. This model provides a phenomenological treatment of hadronic and nuclear interactions within a Reggeon field theory framework, and includes soft and semi-hard parton processes within the ``semi-hard pomeron'' approach.
   \QGSJET was also developed for the simulation of cosmic rays.
   \QGSJET and \EPOS calculations do not rely on the standard PDFs as used in the \PYTHIA generators.

\end{itemize}

The \PYTHIAatwo, \PYTHIA 6 \textsc{AMBT2B} and \EPOSLHC models were used to generate event samples which were processed by the \GEANT{4}-based~\cite{Agostinelli:2002hh} ATLAS simulation framework~\cite{Aad:2010ah}. The simulation also takes into account inactive and inefficient regions of the ATLAS detector. The resulting datasets were used to derive corrections for detector effects and to evaluate systematic uncertainties.

Comparisons to the data corrected to particle level are made with generated events using the \PYTHIAatwo and \textsc{Monash} tunes, the \EPOSLHC tune, and the default \QGSJET tune. These comparisons are shown in \Sect{\ref{sec:results}}.

\section{Event selection}
\label{sec:event}

A dedicated LHC $pp$ run was used for which the average number of $pp$ interactions per bunch crossing, $\langle \mu \rangle$, was low ($0.0028 < \mu < 0.004$).
The maximum instantaneous luminosity was approximately
\SI{1.8e28}{\per\cm\squared\per\s}.
Events were selected for which all subcomponents of the ID were operational and the solenoid magnet was on.
Only events from colliding proton bunches in which the MBTS trigger recorded one or more modules above threshold on either side were considered.
The MBTS trigger efficiency is described in detail in \Sect{\ref{sec:trigeff}}.

The following event selection criteria were applied:

\begin{itemize}
   \item
   A primary vertex with at least two associated tracks constrained to the luminous $z$-region of the measured beam position (termed \emph{beam spot}) was required.
The tracks were required to possess $\pT > 100 \MeV$ and their transverse distance of closest approach to the beam spot ($\dzerobs$) was restricted such that $|\dzerobs| < \SI{4}{\mm}$.
   \item Events were rejected if they had at least one additional vertex with four or more associated tracks.
   Following this selection, the estimated fraction of remaining pile-up events with more than one $pp$ interaction, based on $\langle \mu \rangle$, was about \SI{0.002}{\percent}.
Events containing additional vertices with less than four tracks are dominated by split vertices, where the vertex reconstruction algorithm wrongly reconstructs two vertices from tracks which actually originate from a single vertex~\cite{ATLAS:2010lca}, and by secondary interactions being reconstructed as another primary vertex.
The fraction of events with split vertices or secondary interactions which are rejected by this criterion was estimated from simulation to be \SI{0.01}{\percent}, which is negligible and therefore ignored.
   \item Depending on the phase space under study, additional selections were made on track multiplicity given the required minimum transverse momentum possessed by a track. A minimum number of selected tracks $\nsel \geq 2$ with transverse momentum $\pT > 100 \MeV$, or $\nsel \geq 1$ with $\pT > 500 \MeV$, which satisfy the constraints given in \Sect{\ref{sec:track}}, was required.
\end{itemize}

Following the application of the above selections, the event yield is $9.2 \times 10^6$ for the most inclusive phase space at $\nsel \geq 2$ and $\pT > 100 \MeV$. The phase space with the lowest number of events ($\sim 6.4 \times 10^4$) corresponds to  $\nsel \geq 50$ and $\pT > 500 \MeV$.

\section{Track reconstruction and selection}
\label{sec:track}

Tracks were reconstructed using two approaches as in previous studies at $\sqrt{s} = 7 \TeV$~\cite{Aad:2010ac}. Firstly, an \emph{inside-out} algorithm, starting the pattern recognition from clusters in the Pixel detector, was employed. An additional algorithm with relaxed requirements on the number of silicon hits was employed to reconstruct low-momentum tracks from hits which were unused in the first approach.
This latter method increases the overall efficiency of finding low-\pT tracks (mostly $ 100 < \pT < 400 \MeV$) by up to a factor of two.

\begin{figure}[htbp]
\centering
    \subfloat[\label{fig:raw_eta}]
    {\includegraphics[width=0.48\textwidth]{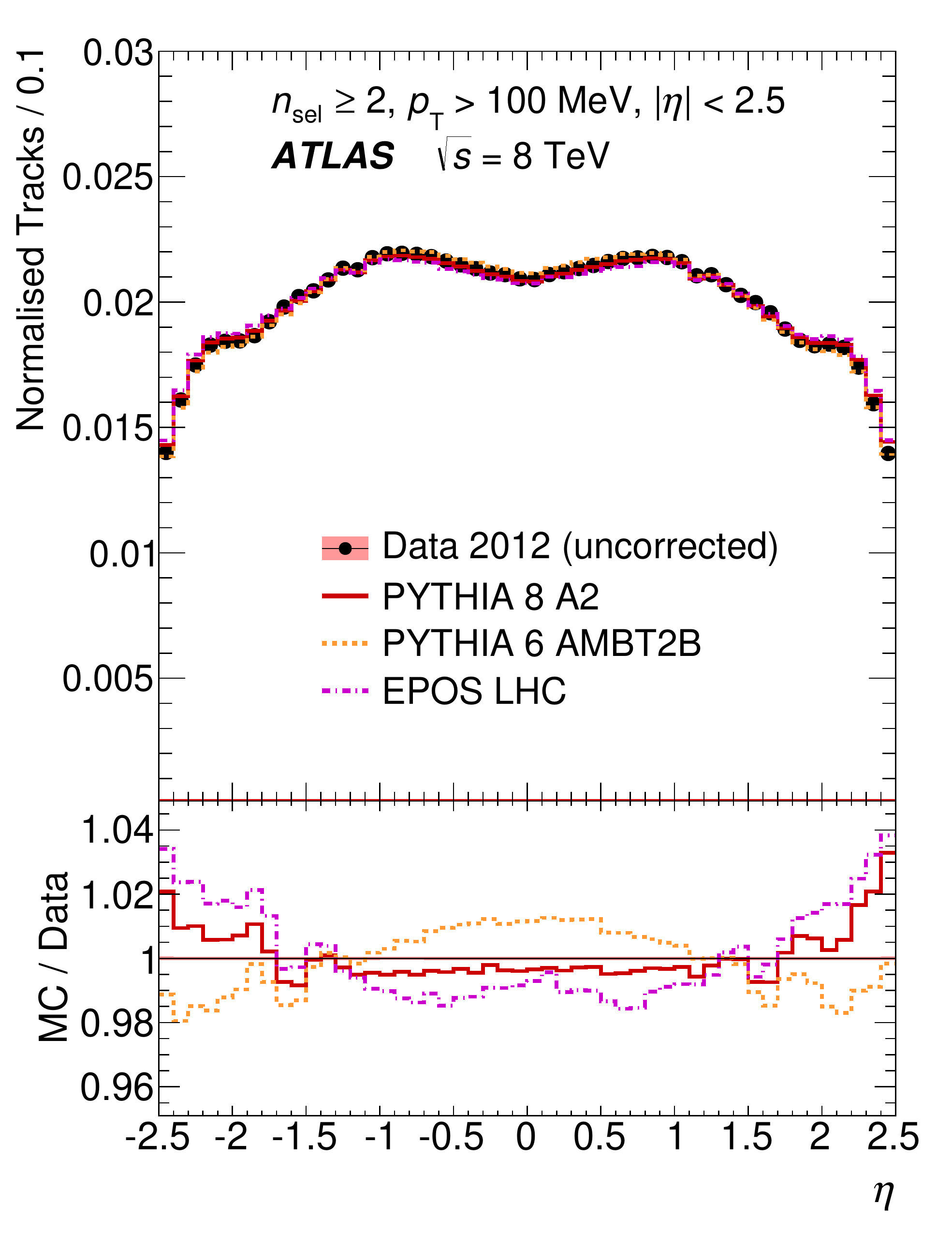}}
    \subfloat[\label{fig:raw_pt}]
    {\includegraphics[width=0.48\textwidth]{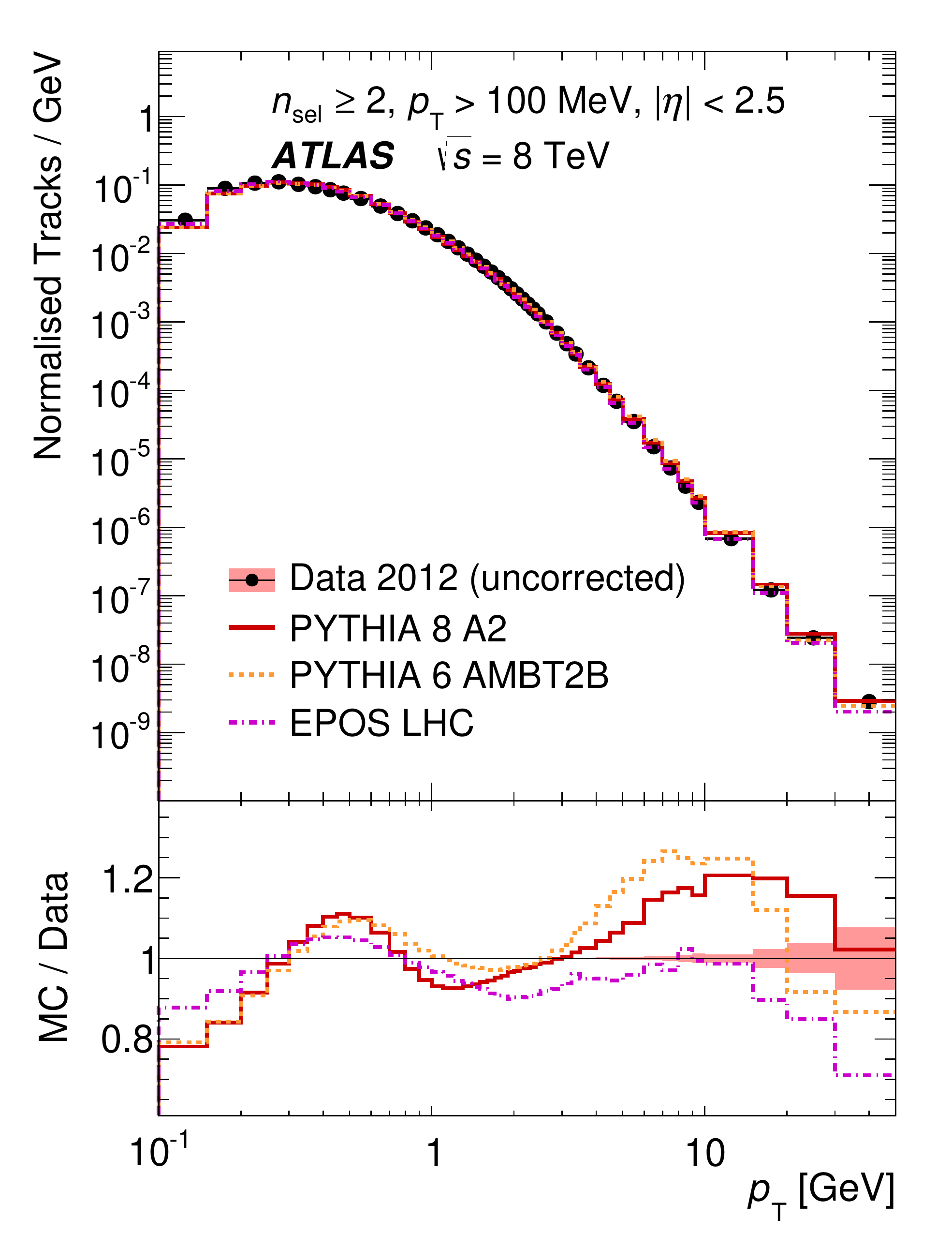}}
    \caption{Distribution of the fraction of selected tracks as a function of (a) pseudorapidity, $\eta$, and (b) transverse momentum, \pT. The predictions of MC models following detector simulation are compared to the data. Bin entries are scaled by the inverse bin width and the resulting distributions are normalised to unity.
    \label{fig:raw}}
\end{figure}

To ensure that well-reconstructed tracks were used at this step, the pseudorapidity and transverse momentum must satisfy $|\eta| < 2.5$ and $\pT > 100 \MeV$. A number of further quality criteria were also applied.
The track must have at least one hit in the pixel detector. A hit in the innermost layer of the pixel detector was required should the extrapolated track have passed through an active region in that layer. At least two, four or six SCT hits are required to be associated with a track for $100 < \pT \leq 200 \MeV$, $200 < \pT \leq 300 \MeV$, or $\pT > 300 \MeV$, respectively. The SCT hit requirements are relaxed in the event of a track trajectory passing through inactive SCT modules.
The probability of the track hypothesis being correct, estimated using the track fit ${\chi}^2$ and $n_{\mathrm{dof}}$, was required to be greater than 0.01 for $\pT > 10 \GeV$ in order to remove tracks with a mis-measured high \pT due to interactions with the material or combinatorial fake high-\pT tracks. 
The distance of closest approach in the transverse ($|\dzeropv|$) and the longitudinal plane ($|\zzeropvsint|$) was also required to be less than \SI{1.5}{\mm} with respect to the primary vertex. These constraints reduce the total fraction of non-primary tracks in the data from around \SI{6}{\percent} to \SI{2}{\percent} (see \Sect{\ref{sec:bg}}). The 
average efficiency to reconstruct a track above $\pT>100 \MeV$ is approximately \SI{70}{\percent}. The efficiency of the two impact parameter requirements 
is around \SI{94}{\percent}, i.e.~applied together they remove approximately \SI{6}{\percent} of all tracks that would pass the other track selection criteria.

\begin{figure}[htbp]
\centering
    {\includegraphics[width=0.48\textwidth]{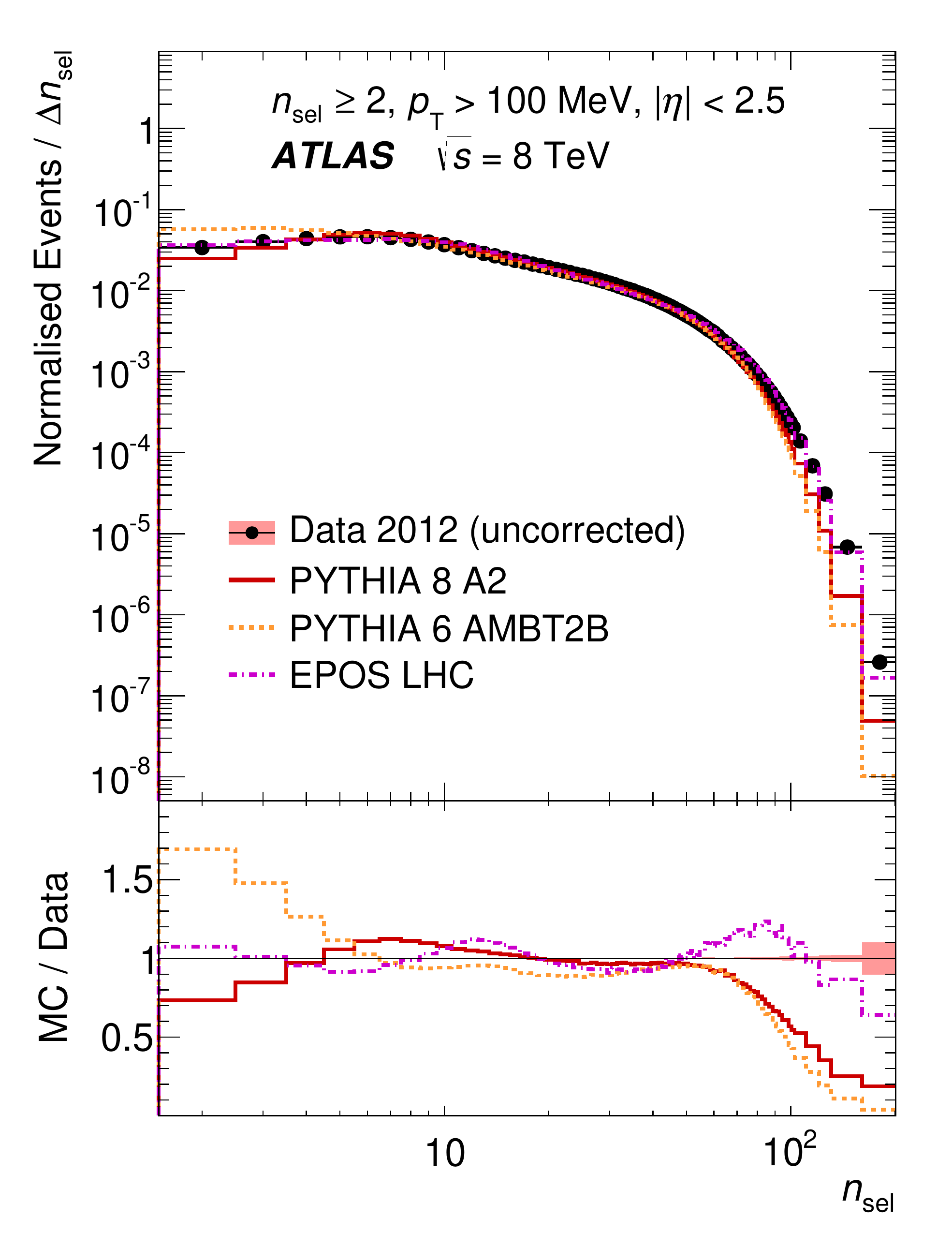}}
    \caption{The fraction of selected events as a function of the track multiplicity \nsel per event. The predictions of MC models following detector simulation are compared to the data. Bin entries are scaled by the inverse bin width and the resulting distributions are normalised to unity.
    \label{fig:raw_nch}}
\end{figure}

\Fig{\ref{fig:raw_eta}} shows the normalised distribution of all selected tracks as a function of pseudorapidity in
 the most inclusive phase space. The models reproduce the data well with discrepancies in $\eta$ at a level of up to \SI{3}{\percent}, which stem from the imperfect description of the \pT spectra (\Fig{\ref{fig:raw_pt}}) by the models, where discrepancies of up to $\sim \SI{30}{\percent}$ are visible.
\Fig{\ref{fig:raw_nch}} shows the normalised distribution of the fraction of all selected events as a function
of track multiplicity per event.

The distributions of the average number of hits per reconstructed track in data and MC simulation as a function of pseudorapidity are shown in \Fig{\ref{fig:auxhits}} in \App{\ref{sec:hitsontrack}}, using events selected for the most inclusive phase space.

\section{Backgrounds}
\label{sec:bg}

Background events and tracks can arise from a number of sources, which are described by order of importance.

Corrections were made to the charged-particle spectra to remove the contribution of charged non-primary particles, i.e.~those not originating from the $pp$ collision.
Non-primary particles are mainly due to hadronic interactions, photon conversions and weak decays.
MC simulations of the shape of the \dzeropv distributions were used to quantify the fractions of primaries, non-primaries from electrons and other non-primaries which satisfy the track quality criteria.
Using the same method as in \Ref{\cite{Aad:2010ac}}, fits were made to the data using \dzeropv distribution templates, taken from the MC simulation, to assess the fractions of the different classes of charged particles.
The fitted impact parameter distributions are shown in \Fig{\ref{fig:auxd0z0}} in \App{\ref{sec:app_ip}}.
The total non-primary fraction was about \SI{3}{\percent} in the $100 < \pT < 150 \MeV$ range and about \SI{2}{\percent} at higher \pT values.
The relative contribution of electrons (including those from rare Dalitz decays) to this fraction was about \SI{35}{\percent} at $\pT < 150 \MeV$ and dropped below \SI{15}{\percent} with rising \pT.
Systematic uncertainties due to aspects of the template fitting method as well as the choice of MC models were added in quadrature.

In contrast to previous measurements at lower energies~\cite{Aad:2010ac,Aad:2010rd}, and in line with the $13 \TeV$ measurement~\cite{Aad:2016mok}, charged particles with a mean lifetime $30 < \tau < \SI{300}{\ps}$ (mostly charged strange baryons) are considered to be non-stable.
The reconstruction efficiency of these short-lived particles and their decay products is strongly momentum-dependent and close to zero for most particles within the measured kinematic range. However, the predicted fraction of the total generated particles associated with charged strange baryons
varies with \pT as well as between MC models. For example, the fractions predicted by \PYTHIAatwo and \EPOSLHC are \SI{5}{\percent} and \SI{13}{\percent}, respectively for $\pT \sim 5 \GeV$.
To lower the model dependence on the overall track reconstruction efficiency, the contribution of such particles to the distributions under study was excluded from the measurement definition. The residual small contamination of reconstructed tracks, which is less than \SI{0.01}{\percent} in $\eta$ and up to \SI{5}{\percent} at high transverse momentum ($30 < \pT < 50 \GeV$), was estimated from simulation using \EPOSLHC and subtracted, and a systematic uncertainty
was assigned following comparisons of the predictions of different MC models.

Fake tracks are reconstructed either due to detector noise or shared hits from more than one charged particle. These were estimated in simulation to be less than \SI{0.1}{\percent} of all tracks.

Beam-induced background, i.e.~beam--gas interactions and scattering from up-stream collimators, was estimated using unpaired bunches. Beam-induced backgrounds as well as pile-up contamination were reduced to a negligible level by the track-level and event-level criteria described in \Sects{\ref{sec:event}}{\ref{sec:track}}.
The cosmic--ray background was found to be negligible using the techniques in \Ref{\cite{Aad:2010ac}}.

\section{Selection efficiencies}
\label{sec:efficiencies}

In order to obtain inclusive spectra for primary charged particles, the data are corrected from detector level to particle level, using corrections which account for inefficiencies due to trigger selection, vertex and track reconstruction.
The methods used to obtain these efficiencies and their systematic uncertainties are described in the following sections.

\subsection{Trigger efficiency}
\label{sec:trigeff}

The trigger efficiency, \trigeff, was measured from a data sample selected using a random control trigger in coincidence with colliding bunches with a minimum requirement of two Pixel and three SCT measurements.
For this efficiency, the requirement of a reconstructed primary vertex was removed from the selection of events to account for possible correlations between the trigger and vertex reconstruction efficiencies.
The trigger efficiency was therefore parameterised as a function of \nselbs, which is defined as the number of tracks in an event that satisfy all track quality criteria; however, instead of the nominal requirements for the impact parameters $\dzeropv$ and $\zzeropv$, only a constraint on the transverse impact parameter with respect to the beam spot, $|\dzerobs| < \SI{1.8}{\mm}$, was applied in order to minimise correlations between the trigger and vertex efficiency corrections.

The trigger efficiency was calculated as the ratio of events triggered by the control trigger, in which the MBTS trigger also accepted the event, to the total number of triggered events in the control sample.
It was determined separately for the trigger requirement in which the signal in at least one or two of the MBTS modules was above threshold.
In order to maximise the recorded data yield, both triggers were combined such that either of the two triggers was required to trigger the event, in which case the corresponding trigger efficiency was applied;
this was done due to pre-scale factors that evolved differently for each trigger during the run.
The result for the trigger requirement in which the signal in at least one of the MBTS modules was above threshold is presented in \Fig{\ref{fig:trigeff_100}} as a function of \nselbs for the most inclusive phase space.
In the $\pT > 500 \MeV$ phase space, the efficiency was measured to be above \SI{98}{\percent} for $\nselbs = 1$ and it rises more rapidly to \SI{100}{\percent} at higher track multiplicities than in the most inclusive phase space.
The efficiency for the trigger requirement in which a signal above threshold was required in at least two of the MBTS modules is lower by about \SI{4}{\percent} for the ($\nselbs = 2, \ \pT > 100 \MeV$) event category, and about \SI{2}{\percent} lower for events with $\nselbs = 1$ and $\pT > 500 \MeV$. It rises more slowly to \SI{100}{\percent} as a function of \nselbs for both \pT requirements.
These additional results are shown in \Fig{\ref{fig:auxtrig}} in \App{\ref{sec:trigeffplots}}.

The trigger requirement was found to introduce no observable bias in the \pT and $\eta$ distributions of selected tracks beyond the statistical uncertainties of the data recorded with the control trigger.
The systematic uncertainties shown in \Fig{\ref{fig:trigeff_100}} due to beam-induced background and tracks from secondary particles were estimated from differences in the trigger efficiency by varying the impact parameter requirements in the track selection.
The total systematic uncertainty on the trigger efficiency in the $\nselbs \geq 2, \ \pT \geq 100 \MeV$ phase space was \SI{0.7}{\percent} for $\nselbs = 2$, decreasing rapidly at higher track multiplicities.
This uncertainty is very small compared to those from other sources.

\subsection{Vertex reconstruction efficiency}
\label{sec:vrtxeff}

The vertex reconstruction efficiency, \vtxeff, was determined for data and MC simulation from the ratio of selected events which satisfy the trigger requirement and contain a reconstructed vertex to the total number of triggered events.
The expected contribution from beam-induced background events is estimated using the same method as described in \Ref{\cite{Aad:2010ac}} and subtracted before measuring the efficiency.
The vertex reconstruction efficiency was parameterised as a function of \nselbs, using the same track quality criteria with modified impact parameter constraints as for the trigger efficiency.

\begin{figure}[htbp]%
    \centering
    \subfloat[\label{fig:trigeff_100}]
    {\includegraphics[width=0.48\textwidth]{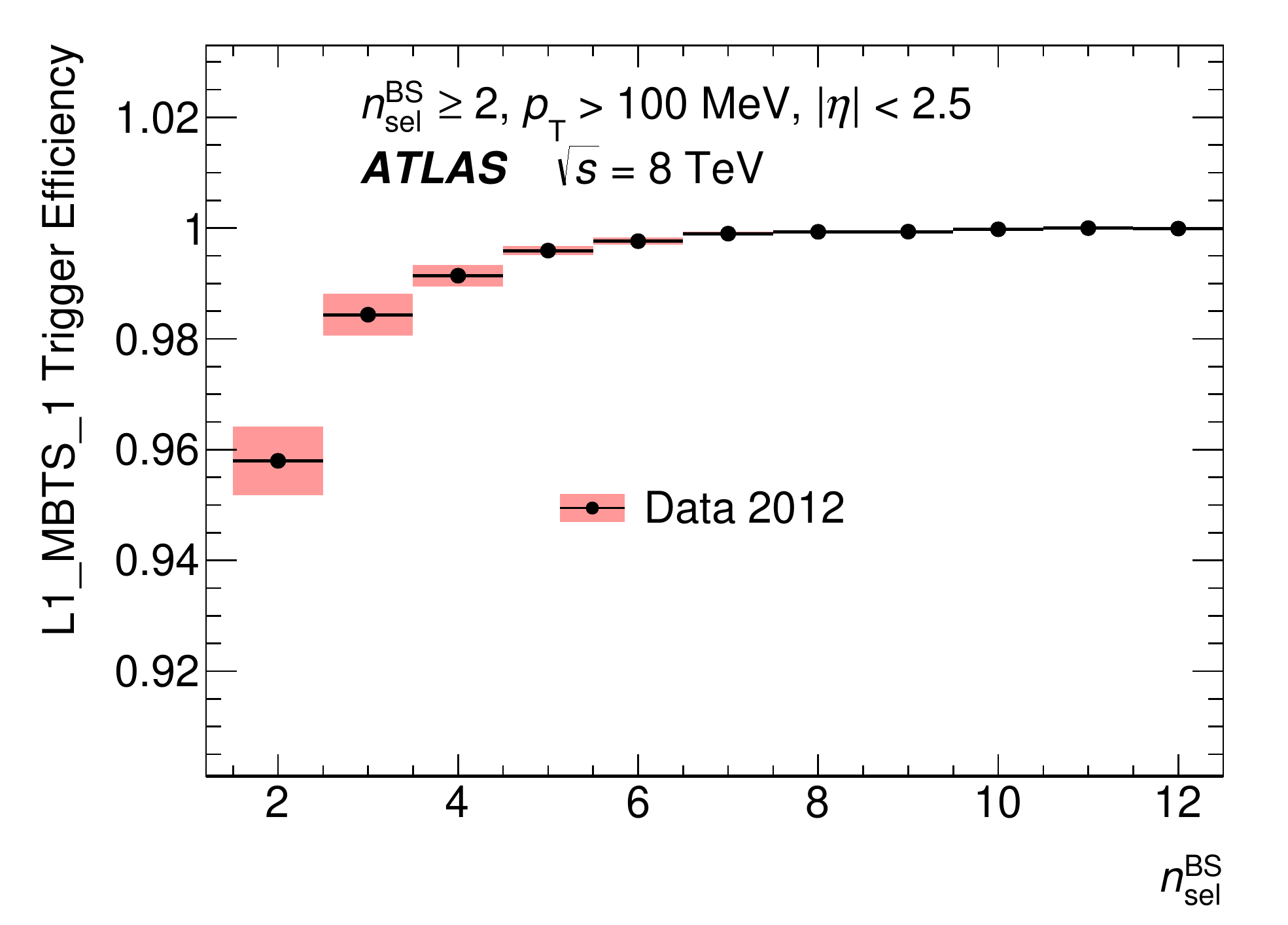}}
    \subfloat[\label{fig:vrtxeff_100}]
    {\includegraphics[width=0.48\textwidth]{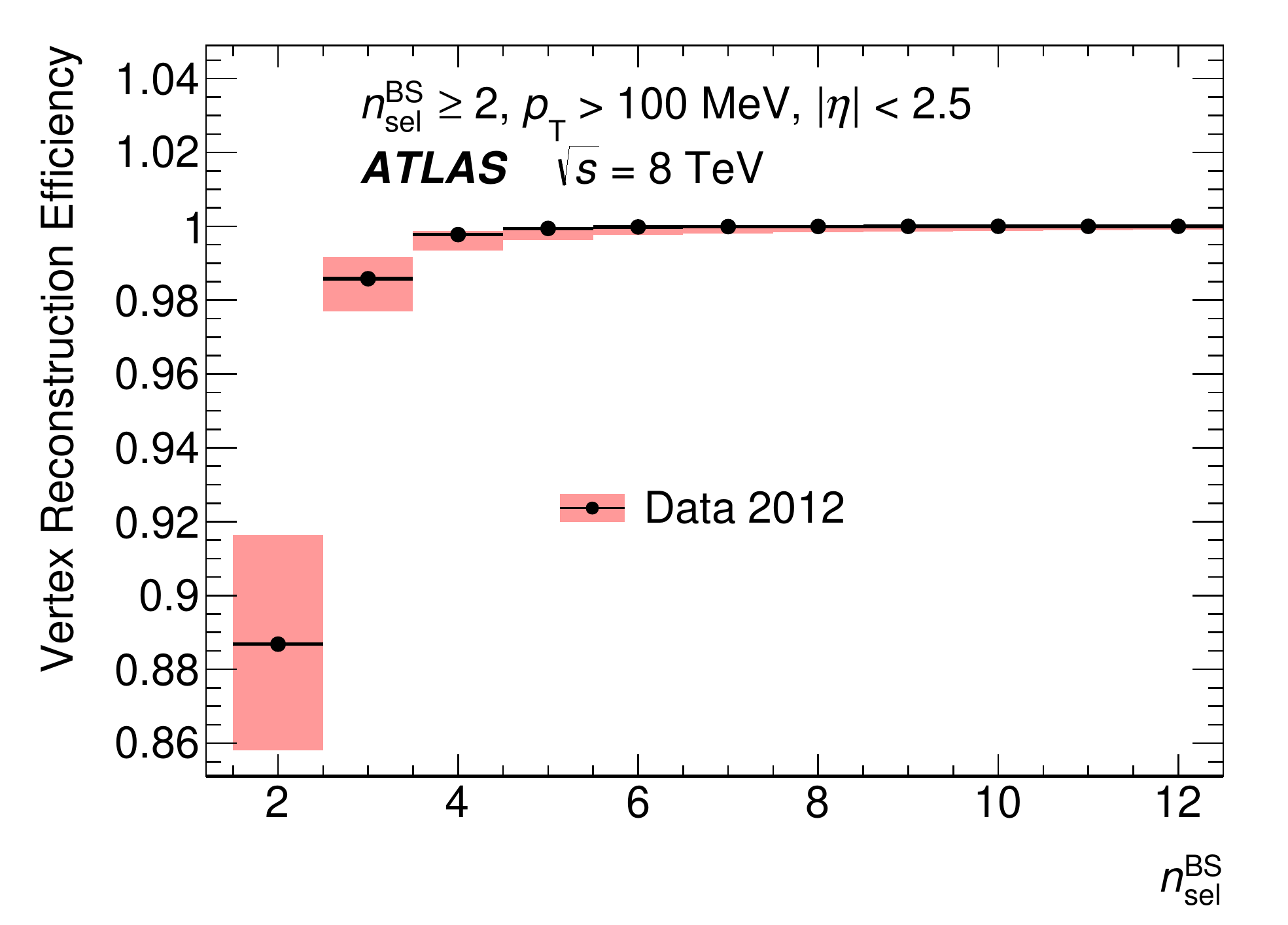}} \\
    \subfloat[\label{fig:trkeff_100_eta}]
    {\includegraphics[width=0.48\textwidth]{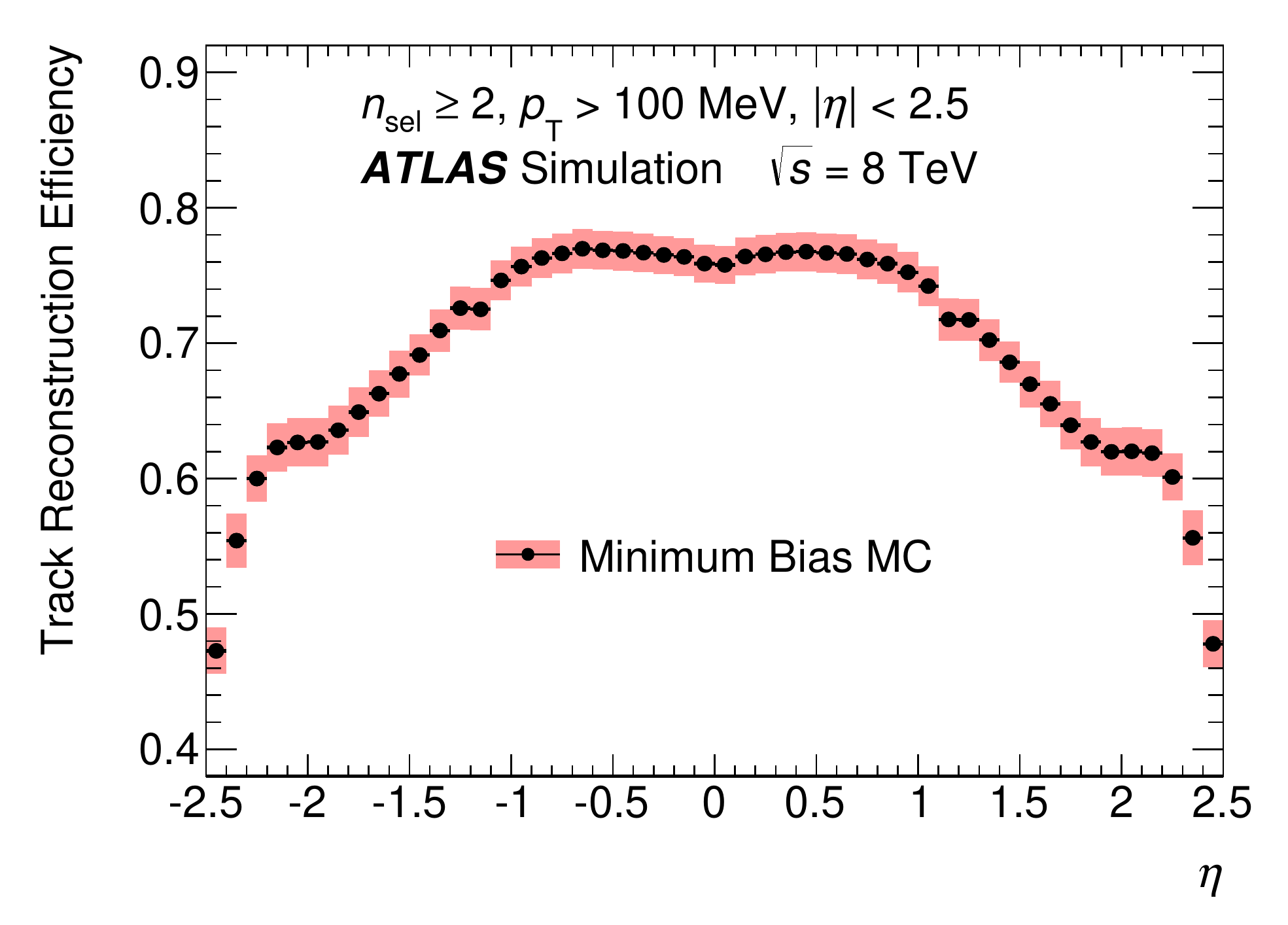}}
    \subfloat[\label{fig:trkeff_100_pt}]
    {\includegraphics[width=0.48\textwidth]{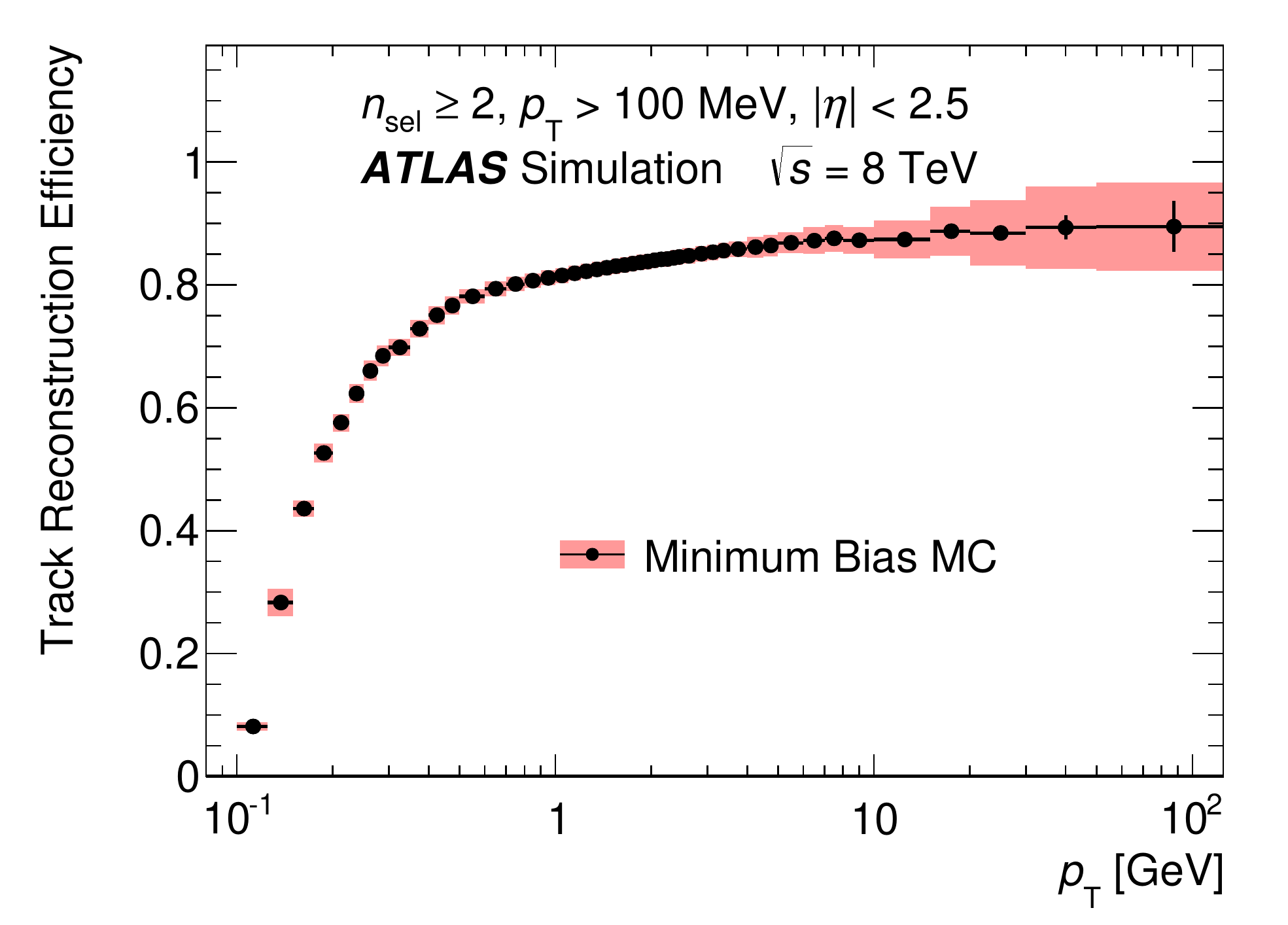}}
    \caption{
    Selection efficiencies for $8 \TeV$ data in the most inclusive measured phase space with transverse momentum $\pT > 100 \MeV$:
(a) The L1\_MBTS\_1 trigger efficiency as a function of the number of selected tracks, \nselbs.
L1\_MBTS\_1 is the requirement that in at least one module of the minimum-bias trigger scintillators a signal above threshold was registered.
(b) The vertex reconstruction efficiency as a function of the number of selected tracks, \nselbs.
(c) The track reconstruction efficiency as a function of the pseudorapidity, $\eta$.
(d) The track reconstruction efficiency as a function of the transverse momentum, \pT.
The shaded areas represent the sum of systematic and statistical errors.
    \label{fig:efficiencies}}
\end{figure}

The result is shown in \Fig{\ref{fig:vrtxeff_100}} as a function of \nselbs for events in the most inclusive phase space with $\pT > 100 \MeV$.
The efficiency was measured to be approximately \SI{89}{\percent} for $\nselbs = 2$, rapidly rising to \SI{100}{\percent} at higher track multiplicities.
For the $\pT > 500 \MeV$ phase space, the result is given in \Fig{\ref{fig:vrtxeff_500}} in \App{\ref{sec:morevtxtrkeffplots}}.
For events with $\nselbs = 2$ in the $\pT > 100 \MeV$ phase space, the efficiency was parameterised as a function of the minimum difference in longitudinal impact parameter ($\Delta z^\mathrm{min}_0$) of track pairs, as well as the minimum transverse momentum (\pTmin) of selected tracks in the event.
For events with $\nselbs = 1$ in the $\pT > 500 \MeV$ phase space, the efficiency was parameterised as a function of $\eta$ of the single track.

The systematic uncertainty was estimated from adding in quadrature the difference between the nominal vertex reconstruction efficiency, measured with beam-induced background removal, and either (1) the vertex reconstruction efficiency measured without beam-induced background removal, or (2) the vertex reconstruction efficiency with a modified impact parameter constraint. The total uncertainty is below \SI{3}{\percent} for $\nselbs = 2$ in the most inclusive phase space, rapidly decreasing at higher track multiplicities. This uncertainty is small compared to those from other sources, except at very low track multiplicities.

\subsection{Track reconstruction efficiency}
\label{sec:trkeff}

The primary track reconstruction efficiency, \trkeff, was determined from MC simulation and parameterised in two-dimensional bins of \pT and $\eta$. This efficiency includes the efficiency of the track selection requirements (see \Sect{\ref{sec:track}}). It is defined as the ratio:

\begin{equation}
\label{eq:track_efficiency}
\trkeff (\pT,\eta) \ = \ \frac
 {\Nmatched (\pTgen,\etagen)}
 {\Ngen (\pTgen,\etagen)} ,
\end{equation}

where \pTgen and \etagen are properties of the generated particle,
$\Nmatched (\pTgen,\etagen)$ is the number of reconstructed tracks matched to a generated primary charged particle in a $(\pTgen,\etagen)$ bin,
and $\Ngen (\pTgen,\etagen)$ is the number of generated primary charged particles in that bin.
A track is considered matched to a generated particle if that particle has the smallest
angular distance $\Delta R$
to the track, if $\Delta R < 0.15$, and if the particle trajectory is compatible with the position of at least one pixel hit of the associated track.

The resulting reconstruction efficiency as a function of $\eta$ integrated over \pT is shown in \Fig{\ref{fig:trkeff_100_eta}} for the most inclusive phase space,
and in \Fig{\ref{fig:trkeff_500_eta}} in \App{\ref{sec:morevtxtrkeffplots}} for the phase space given by $\nsel \geq 1$ and $\pT > 500 \MeV$.
The shape of the $\eta$ distribution is strongly affected by the amount of material traversed by charged particles, in particular the passive material in supporting structures between the Pixel and SCT detector. A larger amount of passive material is located at high $|\eta|$ and increases the probability of particles to undergo particle--matter interactions such as hadronic interactions, which reduces the track reconstruction efficiency.
The approximately constant efficiency at $|\eta| \sim 2.1$ is due to the particles passing through an increasing number of layers in the ID end-cap.
\Fig{\ref{fig:trkeff_100_pt}} shows the efficiency as a function of \pT integrated over $\eta$.
The \pT dependence is largely due to the requirement on the minimum number of silicon hits in the track reconstruction algorithms, which is less likely to be fulfilled by lower-\pT tracks.

As the track reconstruction efficiency is determined from MC simulation, its systematic uncertainties result from model dependencies and from the uncertainty of the detector material description used in the simulation.
Since the generated particle composition and the reconstructed track composition differs between MC tunes, a small model-dependence of the track reconstruction efficiency can be observed, leading to an additional systematic uncertainty due to the particle composition.
The impact of the choice of physics models for hadronic interactions in \GEANT{4} simulation is also taken into account.

The amount of material within the ID was constrained to within $\pm \SI{5}{\percent}$, based on extensive studies of material interactions~\cite{Lukas:1956718}.
The systematic uncertainties on the track reconstruction efficiencies were obtained by comparing the predictions of simulations which assume the nominal ID material distribution with two special simulations in which the assumed material was varied. For one simulation, the amount of non-sensitive ID material was increased by \SI{5}{\percent} in terms of radiation length $X_0$. In the other, the Pixel service material was increased by \SI{10}{\percent} in $X_0$.
These studies give rise to an average systematic uncertainty on the track reconstruction efficiency of \SIrange[range-phrase = --]{1.6}{1.7}{\percent}
in the central region and up to \SI{3.5}{\percent} in the forward region, with larger uncertainties up to \SI{8}{\percent} for particles with very low transverse momenta of $\pT < 150 \MeV$.
This is the dominant contribution to the total systematic uncertainty in most regions of the measured distributions.
The reduction of this uncertainty with respect to measurements at $\sqrt{s} = 7 \TeV$, due to our improved knowledge of the ID material distribution, is about \SI{50}{\percent} in the central region and rises to as much as \SI{65}{\percent} in the forward region.

Systematic uncertainties due to simulation of the efficiency of the requirements on the number of hits
associated with a track, the impact parameter requirements, and the efficiency of the track-fit $\chi^2$ probability requirement were found by comparing each selection efficiency in data and MC simulation. The sum in quadrature of these uncertainties varies between \SI{0.5}{\percent} and \SI{1.6}{\percent} for all $\eta$ values and $\pT < 10 \GeV$, and increases to as much as \SI{8}{\percent} for high-momentum tracks above $\pT > 30 \GeV$ in the most forward regions.

The systematic uncertainty due to different fractions of positively and negatively charged tracks in data and MC simulation was found to be negligible.

The total uncertainty of the track reconstruction efficiency, shown in \Figs{\ref{fig:trkeff_100_eta}}{\ref{fig:trkeff_100_pt}}, was obtained by adding all effects in quadrature and is dominated by the uncertainty from the material description.

\section{Correction procedure}
\label{sec:corrections}

In order to obtain inclusive particle-level distributions, all measured
detector-level distributions were corrected by an event-by-event weight, and track distributions were additionally corrected by a track-by-track weight, to compensate for the inefficiencies of the data selection and the reconstruction algorithms, as well as for contaminations due to various sources of background.
Furthermore, a Bayesian unfolding procedure~\cite{D'Agostini:1994zf} was applied to compensate for migration and resolution effects in the observed multiplicity and transverse momentum distributions.

\subsection{Event and track weights}
\label{sec:eventandtrackweights}

All selected events were corrected with an event-by-event weight to compensate for the inefficiencies of the MBTS trigger selection and the vertex reconstruction algorithm.
The total event weight $w_\mathrm{ev}$ is parameterised as:

\begin{equation}
\label{eq:event_weight}
w_\mathrm{ev}(\nselbs,x) = \frac
{1}
{{\trigeff(\nselbs)} \cdot {\vtxeff(\nselbs,x)}} .
\end{equation}

The parameter $x$ represents a combination of \pTmin for all selected tracks, the minimum difference in longitudinal impact parameter ($\Delta z^\mathrm{min}_0$) for track pairs, and $\eta$ of a single track (for events with only one selected track), as described in \Sect{\ref{sec:vrtxeff}}.
In addition, the MC simulation events were weighted such that the vertex $z$-distribution agrees with that observed in data.

Furthermore, a track-by-track weight, $w_\mathrm{trk}(\pT,\eta)$, was estimated for each selected track as a function of the transverse momentum and pseudorapidity assigned to the track, based on
the track reconstruction efficiency, $\trkeff(\pT,\eta)$,
the fraction of non-primary tracks, $\fnonp(\pT,\eta)$,
the fraction of tracks associated with a strange baryon, $\fsb(\pT,\eta)$,
and the fraction of additional tracks corresponding to particles outside the kinematic range but migrating into the kinematic region due to resolution effects, $\fokr(\pT,\eta)$:

\begin{equation}
\label{eq:track_weight}
w_\mathrm{trk}(\pT,\eta) = \frac
 {1 - \fnonp(\pT,\eta) - \fsb(\pT,\eta)  - \fokr(\pT,\eta)}
 {\trkeff(\pT,\eta)} .
\end{equation}

The quantities $\trkeff(\pT,\eta)$, $\fsb(\pT,\eta)$ and $\fokr(\pT,\eta)$ were evaluated using MC models.
The quantification of the contamination due to non-primary tracks and strange baryons is described in \Sect{\ref{sec:bg}}.

\subsection{Correction to \dnch}

Only the event-level corrections for the trigger and vertex inefficiencies 
were applied to the charged-particle multiplicity distribution. 
Thereafter, a Bayesian unfolding process was applied to correct 
the observed multiplicity \nsel to the true number of primary 
charged particles \nch. This is the same procedure as was applied in \Ref{\cite{Aad:2010ac}}, using five iterations in the $\pT > 100 \MeV$ phase space, and four iterations for $\pT > 500 \MeV$. After the unfolding, a correction was made to the resulting primary-charged-particle multiplicity distribution to account for events migrating out of the 
multiplicity range required by the phase space.

The corrected distribution \dnch was integrated over \nch to give the total number of events $\Nev$. The quantity $\Nev$ was then used to normalise the distributions \dNdpT and \dNdeta,
as well as the multiplicity distribution itself, \dnchnorm.

\subsection{Correction to \dNdpT and \dNdeta}

Corrections were made for trigger requirements, vertex and track reconstruction inefficiencies, migration effects due to the resolution of reconstructed track parameters, and the influence of non-primary tracks. A Bayesian unfolding method, similar to that used to correct the \nch spectra, was then employed to give the \dNdpT distribution, using four iterations in the $\pT > 100 \MeV$ phase space, and up to five iterations for $\pT > 500 \MeV$.
Fake high-\pT tracks are already suppressed by the $\chi^2$ probability requirement in the track selection, and remaining fake tracks are also unfolded for by this procedure.

\subsection{Correction to \meanpT versus \nch}

The \meanpT versus \nch distribution was evaluated in the following way. Corrections were made to two separate spectra: the distribution of the $\Sigma_{i} \pT(i)$ (where the summation is made over the transverse momentum of all selected tracks in all events within a certain range of track multiplicity) versus the number of selected tracks per event, and the distribution of the sum of all selected tracks in all events within a certain range of track multiplicity versus the number of selected tracks per event.
The distributions were first corrected with the appropriate track weights, which was followed by Bayesian unfolding. Finally, the ratio of the two spectra was taken to obtain the corrected \meanpT versus \nch distribution.

\section{Systematic uncertainties}
\label{sec:systematics}

In the analysis procedure, most of the individual sources of systematic uncertainty given below were applied separately as variations of the event or track weights, producing new distributions which were used to obtain alternative versions of the final corrected and unfolded results.
Other sources were assessed by varying the input distributions (e.g.~in \nch distributions, the multiplicity of each event was randomly varied with probabilities corresponding to the uncertainties on the track reconstruction efficiencies) or unfolding matrices (using statistical variations, or matrices obtained from different MC generators) which were used for the Bayesian unfolding procedure, thus producing the alternative results.
In all these cases, the differences from the nominal distributions were taken as systematic uncertainties.

The following sources of systematic uncertainty in the corrected distributions were considered.

\begin{itemize}

   \item Incomplete knowledge of the material distribution in the ID affects the measured spectra by between \SI{1}{\percent} and \SI{8}{\percent}. This source of systematic uncertainty is described in detail in \Sect{\ref{sec:trkeff}}. The total uncertainty due to the material distribution is typically less than \SI{5}{\percent} over all distributions other than at $\pT < 150 \MeV$ in the transverse momentum spectrum, at $\nch \geq 120$ in the multiplicity spectrum of the $\pT > 100 \MeV$ phase space, and at $\nch \geq 70$ in the multiplicity spectrum of the $\pT > 500 \MeV$ phase space.
   This is the dominant uncertainty on \dNdeta, and the leading or next-to-leading uncertainty in all other distributions.

   \item Different \pT spectra in the MC models and data lead to differences of up to \SI{2}{\percent} in the average track reconstruction efficiency per \nch interval. For the final \dnch spectra in the most inclusive phase space, this effect becomes as large as \SI{12}{\percent} at the highest multiplicities.

   \item The relative uncertainty on the fraction of non-primary tracks is \SI{15}{\percent}, while the relative uncertainty on the fraction of reconstructed strange baryons is \SI{50}{\percent}, as described in \Sect{\ref{sec:bg}}. The total uncertainty of both sources in the corrected distributions is \SI{3.5}{\percent} or smaller and is not a dominant uncertainty in any region.

   \item Different particle types have different reconstruction efficiencies. For example, at $\pT \sim 1 \GeV$ the reconstruction efficiency of charged pions is $\sim \SI{82}{\percent}$, whereas for kaons and protons it is $\sim \SI{80}{\percent}$ and $\sim \SI{75}{\percent}$, respectively. Although the MC generators give consistent efficiencies, the relative fractions of these generated particles vary between the models. For example, \PYTHIAatwo (\EPOSLHC) gives fractions of \SI{77}{\percent} (\SI{72}{\percent}), \SI{14}{\percent} (\SI{18}{\percent}) and \SI{9}{\percent} (\SI{10}{\percent})
for generated pions, kaons and protons, respectively, at $\pT \sim 1 \GeV$.
Differences in particle composition therefore lead to an uncertainty on the overall track reconstruction efficiency, which varies between \SI{0.2}{\percent} and \SI{1}{\percent} for the corrected distributions. This is not a dominant uncertainty in any region.   

\item Systematic uncertainties on the overall track reconstruction efficiency that
 are associated with the choice of track--particle matching algorithms (\SI{0.4}{\percent}) and the
  choice of physics 
models for MC simulation (\SI{0.3}{\percent}) are also accounted for, and are not a dominant uncertainty in any
 region of the corrected distributions. 

   \item To account for momentum resolution differences between data and MC simulation, which can arise, for example, via imperfect knowledge of the detector alignment, an uncertainty of \SI{5}{\percent} was assigned to tracks with $\pT < 150 \MeV$. At higher values of \pT a one-sided uncertainty of \SI{-7}{\percent} for $10 < \pT < 30 \GeV$ and \SI{-9}{\percent} for $\pT > 30 \GeV$ tracks was assigned, as in the previous work at $\sqrt{s} = 7 \TeV$~\cite{Aad:2010ac}, due to the steeply falling \pT spectrum in combination with the lower momentum resolution in data.
   This is combined with a one-sided uncertainty due to the estimated fraction of mis-measured high-\pT tracks, which increases with transverse momentum to as much as \SI{-16}{\percent} for $\pT > 30 \GeV$ tracks.
   The effect on the corrected distributions is typically negligible, except in the corrected \pT spectra.

   \item Differences in the efficiencies of track quality criteria between data and MC simulation give rise to systematic uncertainties in the final spectra which are typically below \SI{1}{\percent}, except at transverse momenta above $10 \GeV$ and at high multiplicities, reaching as much as \SI{6}{\percent} and \SI{5}{\percent}, respectively. However, this remains a small uncertainty compared to those from other sources in the same regions.

   \item Event-level uncertainties on the trigger efficiency and vertex reconstruction efficiency give rise to systematic uncertainties of up to \SI{3}{\percent} in the lowest multiplicity intervals of the \dnch spectra. However, even in these regions this uncertainty is dominated by other sources.

   \item For each presented distribution, closure tests were performed. A closure test applies the full nominal correction procedure to reconstructed MC simulation events and quantifies the degree to which the generated particle-level distribution is reproduced.

The degree of non-closure is typically less than \SI{1}{\percent} and/or below the level of statistical uncertainties. Larger non-closures were found for the lower end of the \pT spectrum, $100 < \pT < 150 \MeV$, where the non-closure is found to be \SI{6}{\percent} due to momentum resolution effects, and in the low-multiplicity region of the average transverse momentum \meanpT as a function of \nch, with up to \SI{4}{\percent} non-closure in the $\pT > 100 \MeV$ phase space due to assumptions made in the unfolding procedure. All of these non-closures were taken into account as an additional source of systematic uncertainty.

   \item Uncertainties associated with the unfolding technique are estimated as the degree of non-closure following a modified correction procedure, i.e.~obtained in corrected multiplicities after varying the input spectra and unfolding matrix. This is the dominant uncertainty on \dNdpT for transverse momentum values of $\pT > 10 \GeV$, for which the uncertainty varies from \SIrange{6}{20}{\percent}, as well as over the entire range of \meanpT versus \nch. It is also the largest uncertainty in the low and high multiplicity regions of \dnch, for which it has values between \SI{1}{\percent} and \SI{12}{\percent}.

\end{itemize}

All sources of systematic uncertainty are added in quadrature, thus yielding the total systematic uncertainties which are shown as shaded areas in the figures in the next section. 
The total systematic uncertainties in the two most inclusive phase spaces, at $\pT>100 \MeV$ ($\pT > 500 \MeV)$, range 
from \SIrange{1.8}{3.6}{\percent} (\SIrange{1.3}{2.1}{\percent}) in the final \dNdeta distributions, 
from \SIrange{1.6}{30}{\percent} in the final \dNdpT distributions, 
from \SIrange{3}{21}{\percent} (\SIrange{2}{16}{\percent}) in the final \dnch spectra, and 
from \SIrange{1.3}{4}{\percent} (\SIrange{0.5}{2.2}{\percent}) in the final \meanpT versus \nch distributions. 
The lowest uncertainties within these ranges are found at central pseudorapidity ($\eta=0$), around medium transverse momentum values ($\pT \sim 1 \GeV$), and around average multiplicity values of $\nch \sim 20$.

\section{Results}
\label{sec:results}

\begin{figure}[htbp]
\centering
    \subfloat[\label{fig:eta_100_2}]
{\includegraphics[width=0.47\textwidth]{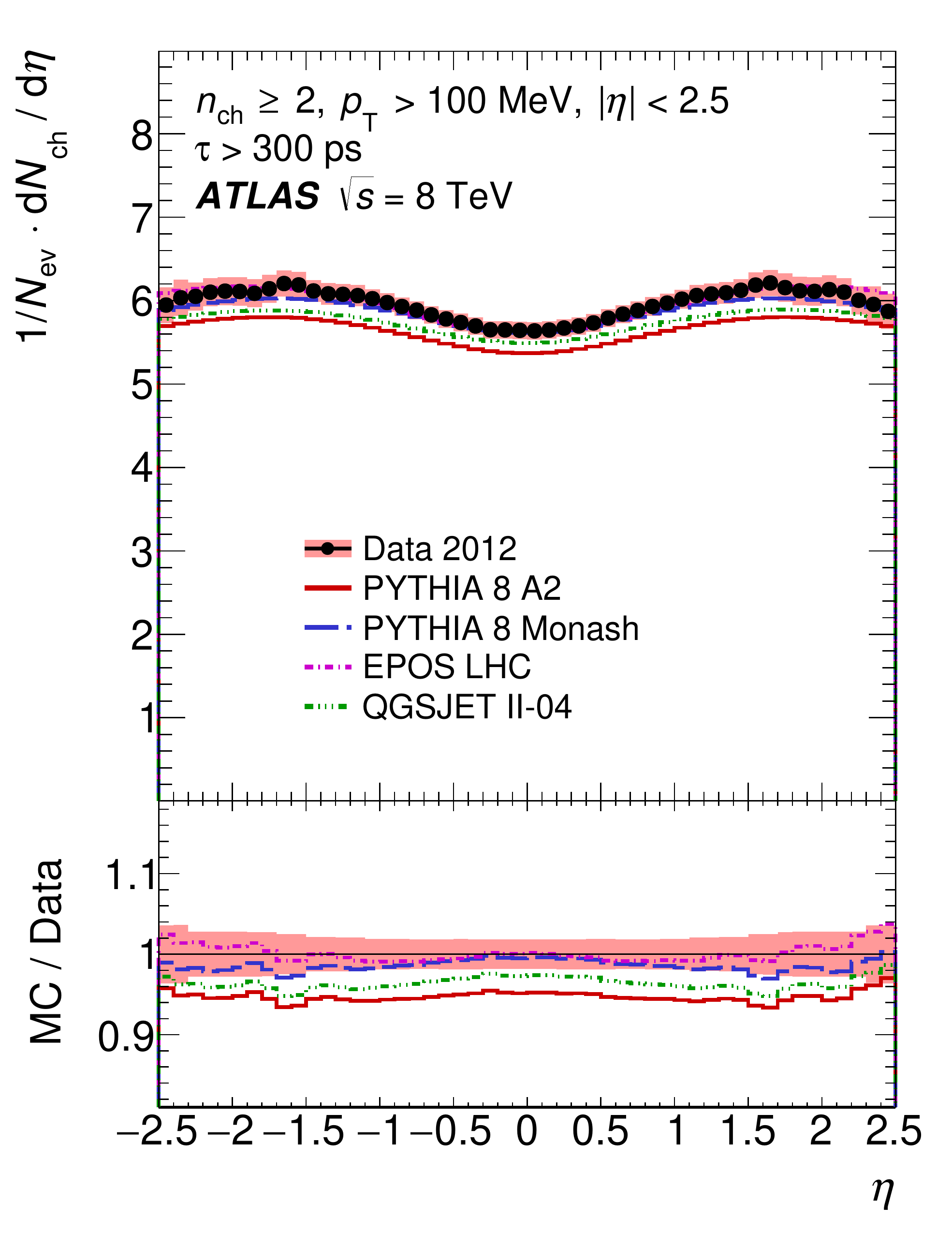}}
    \subfloat[\label{fig:pt_100_2}]
{\includegraphics[width=0.47\textwidth]{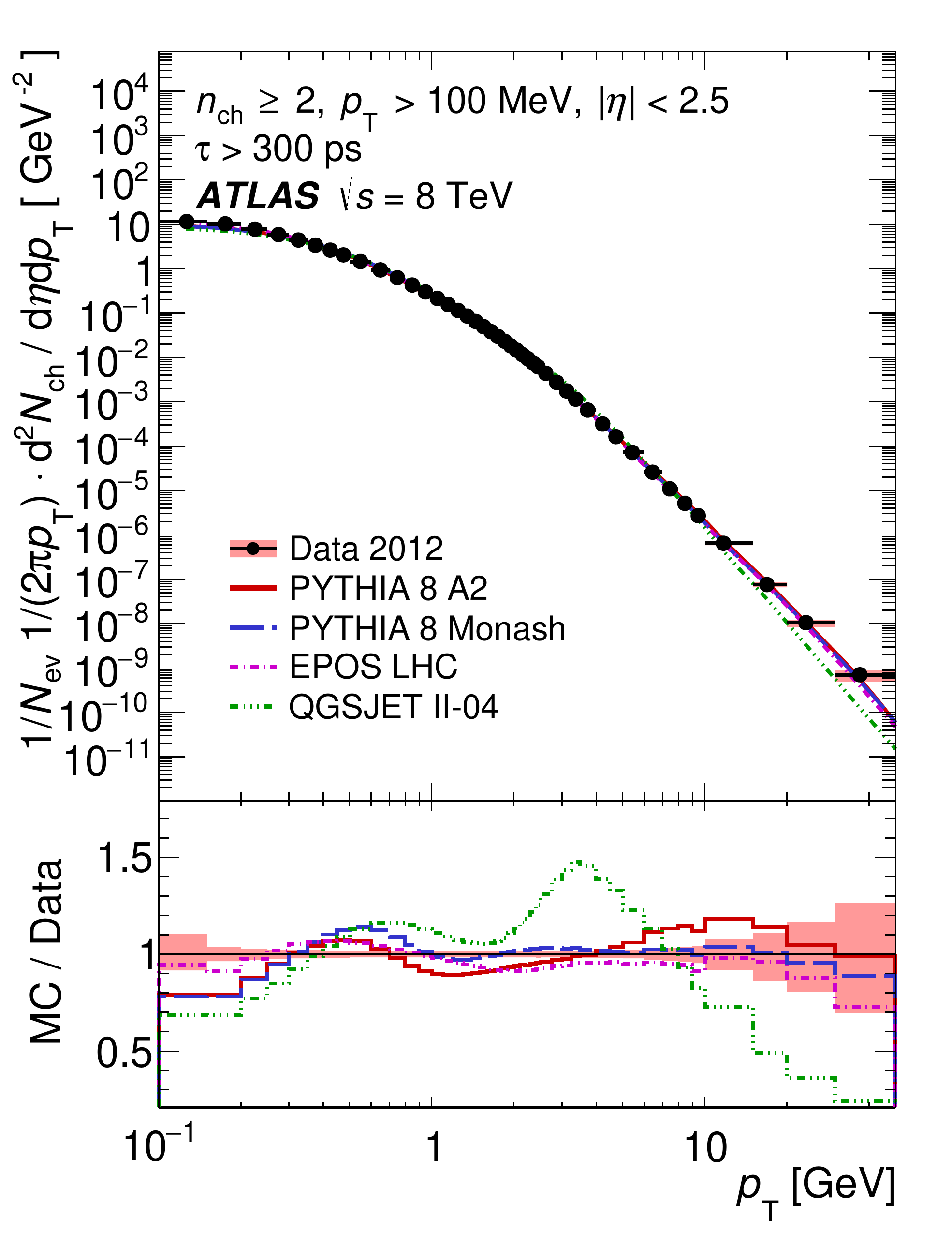}}\\
    \subfloat[\label{fig:nch_100_2}]
{\includegraphics[width=0.47\textwidth]{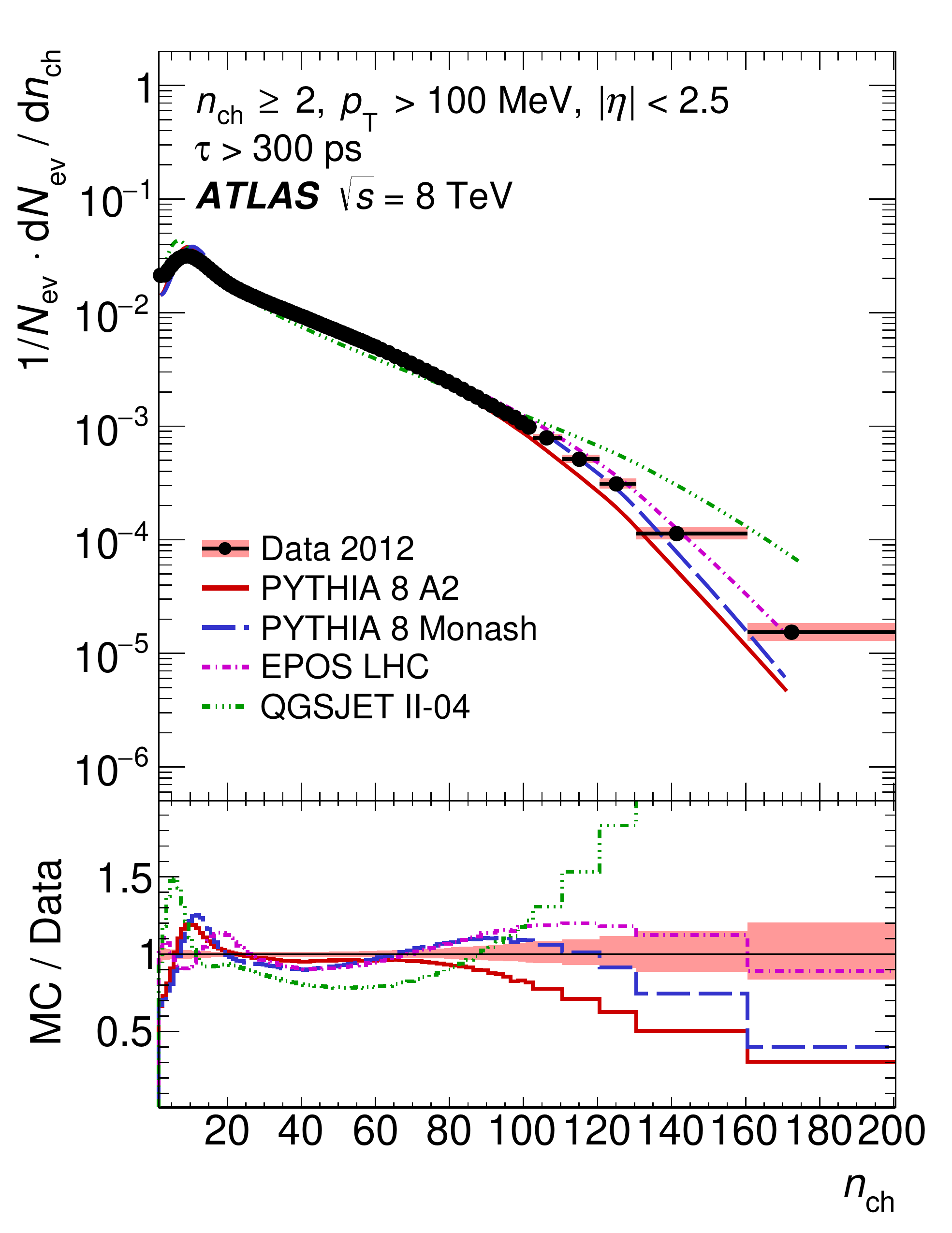}}
    \subfloat[\label{fig:avgpt_100_2}]
{\includegraphics[width=0.47\textwidth]{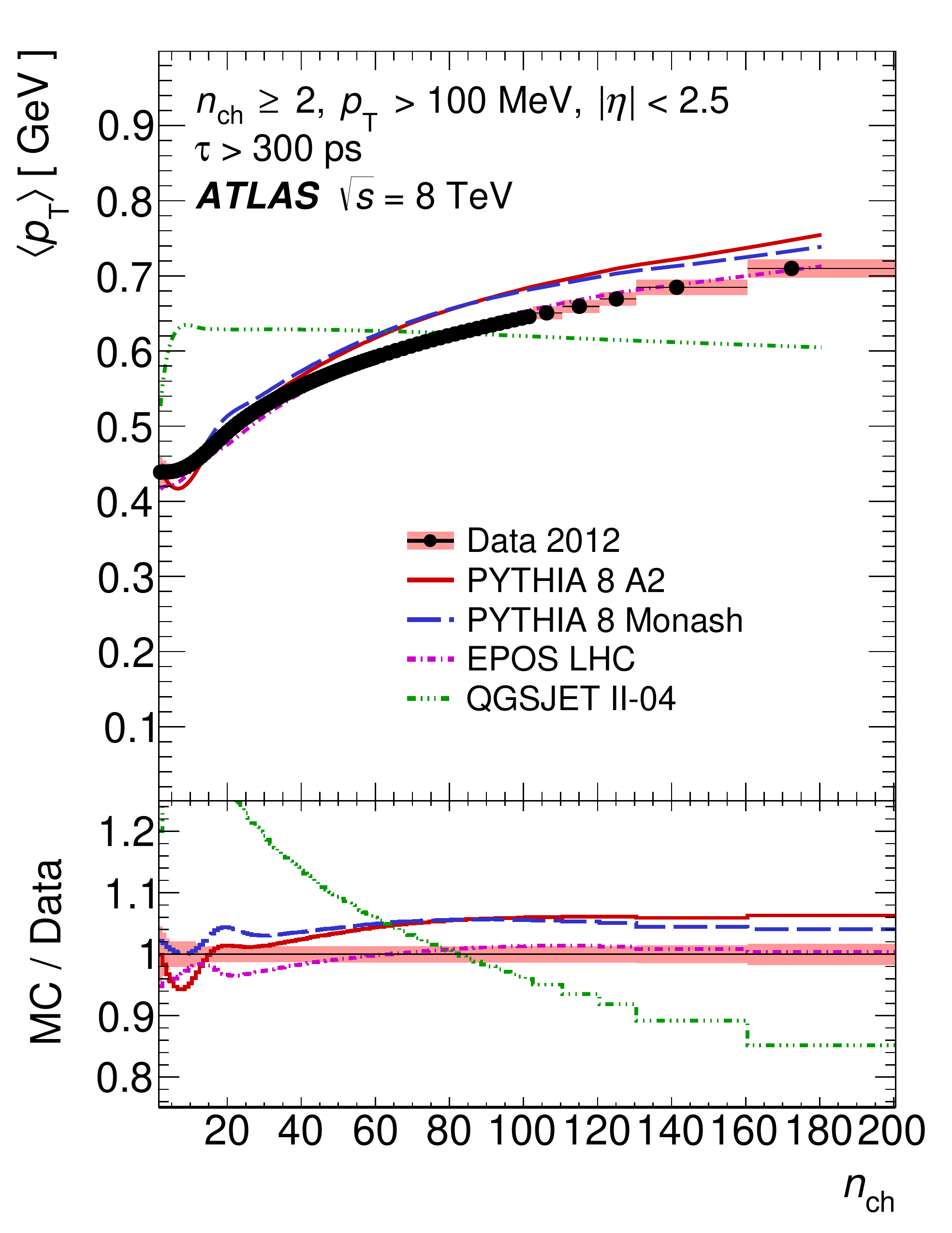}}
    \caption{Distributions of primary charged particles in events for which $\nch \geq 2$, $\pT > 100 \MeV$ and $|\eta| < 2.5$ as a function of (a) pseudorapidity, $\eta$, (b) transverse momentum, \pT, (c) multiplicity, \nch, and (d) average transverse momentum, \meanpT, versus multiplicity.
The data, represented by dots, are compared to various particle-level MC predictions, which are shown by curves.
The shaded areas around the data points represent the total statistical and systematic uncertainties added in quadrature.
\label{fig:final_100_2}}
\end{figure}

\begin{figure}[htbp]
\centering
    \subfloat[\label{fig:eta_500_1}]
{\includegraphics[width=0.47\textwidth]{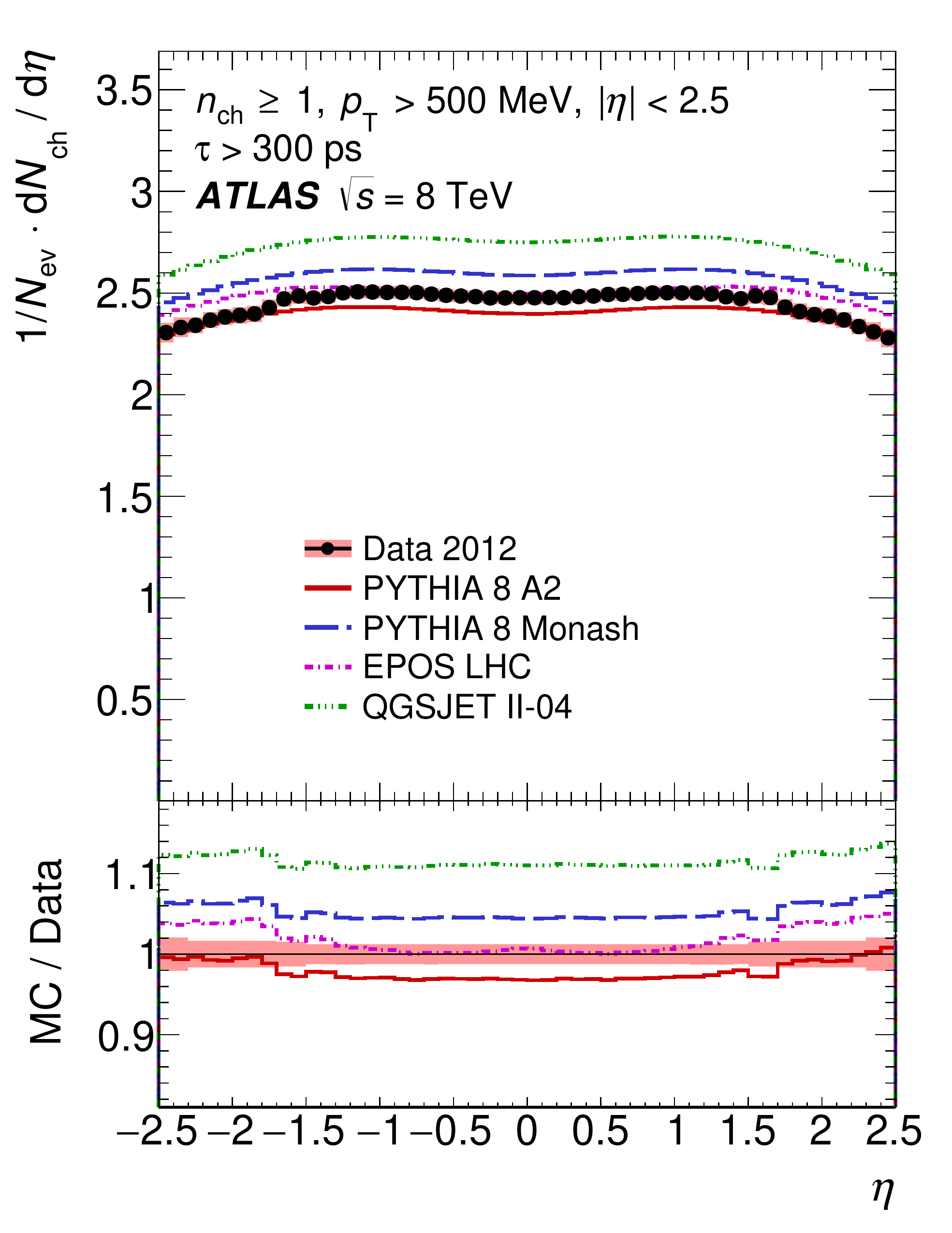}}
    \subfloat[\label{fig:pt_500_1}]
{\includegraphics[width=0.47\textwidth]{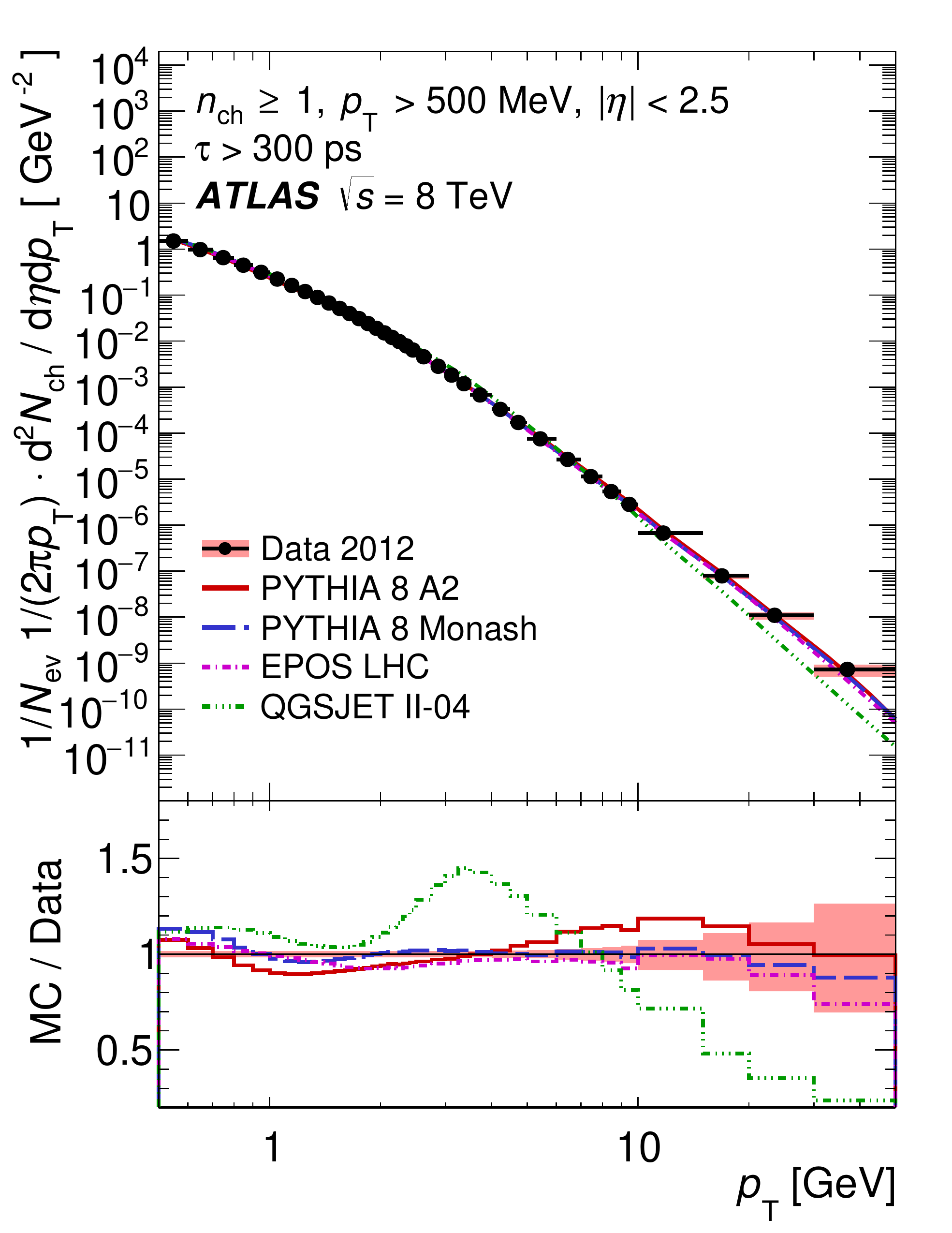}}\\
    \subfloat[\label{fig:nch_500_1}]
{\includegraphics[width=0.47\textwidth]{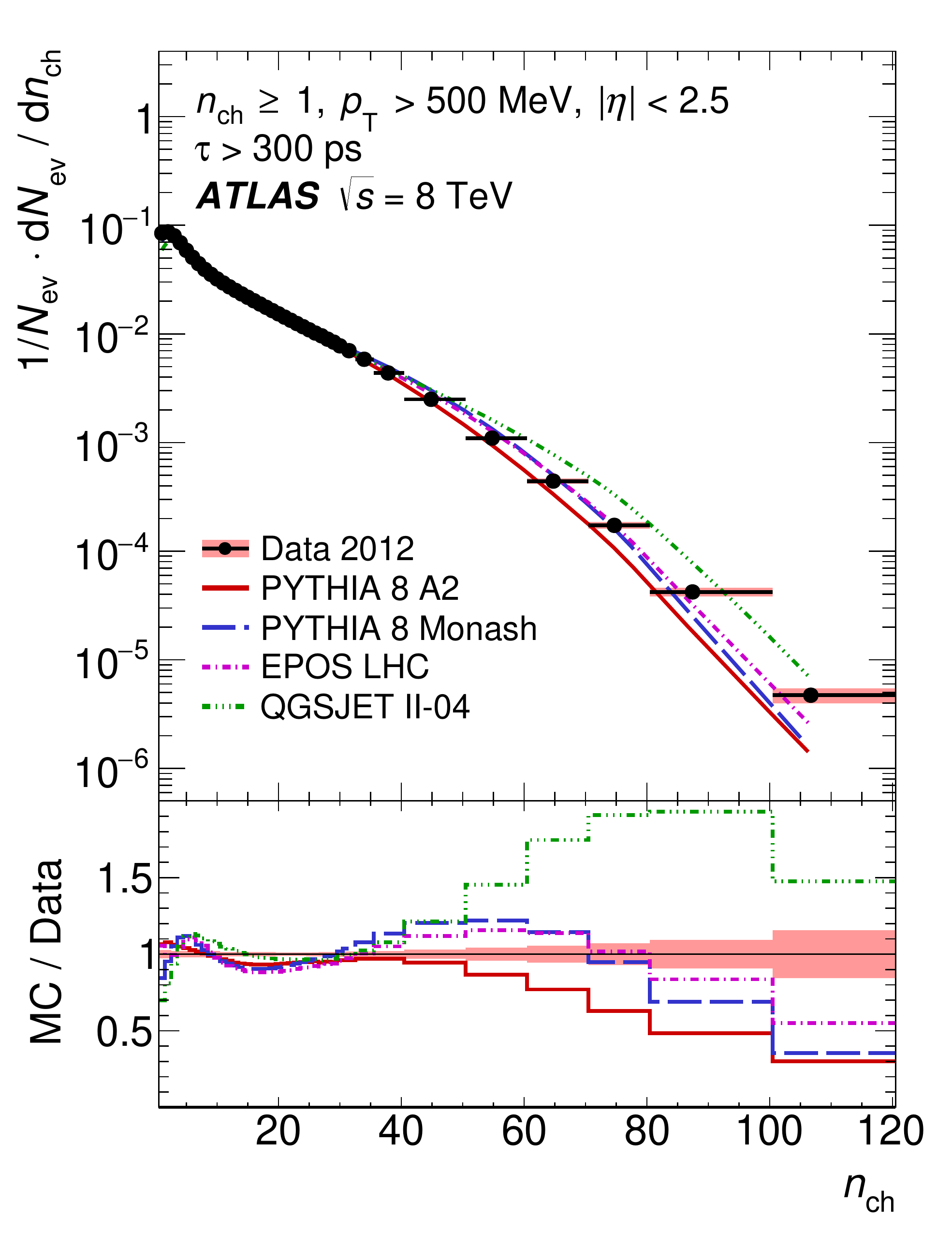}}
    \subfloat[\label{fig:avgpt_500_1}]
{\includegraphics[width=0.47\textwidth]{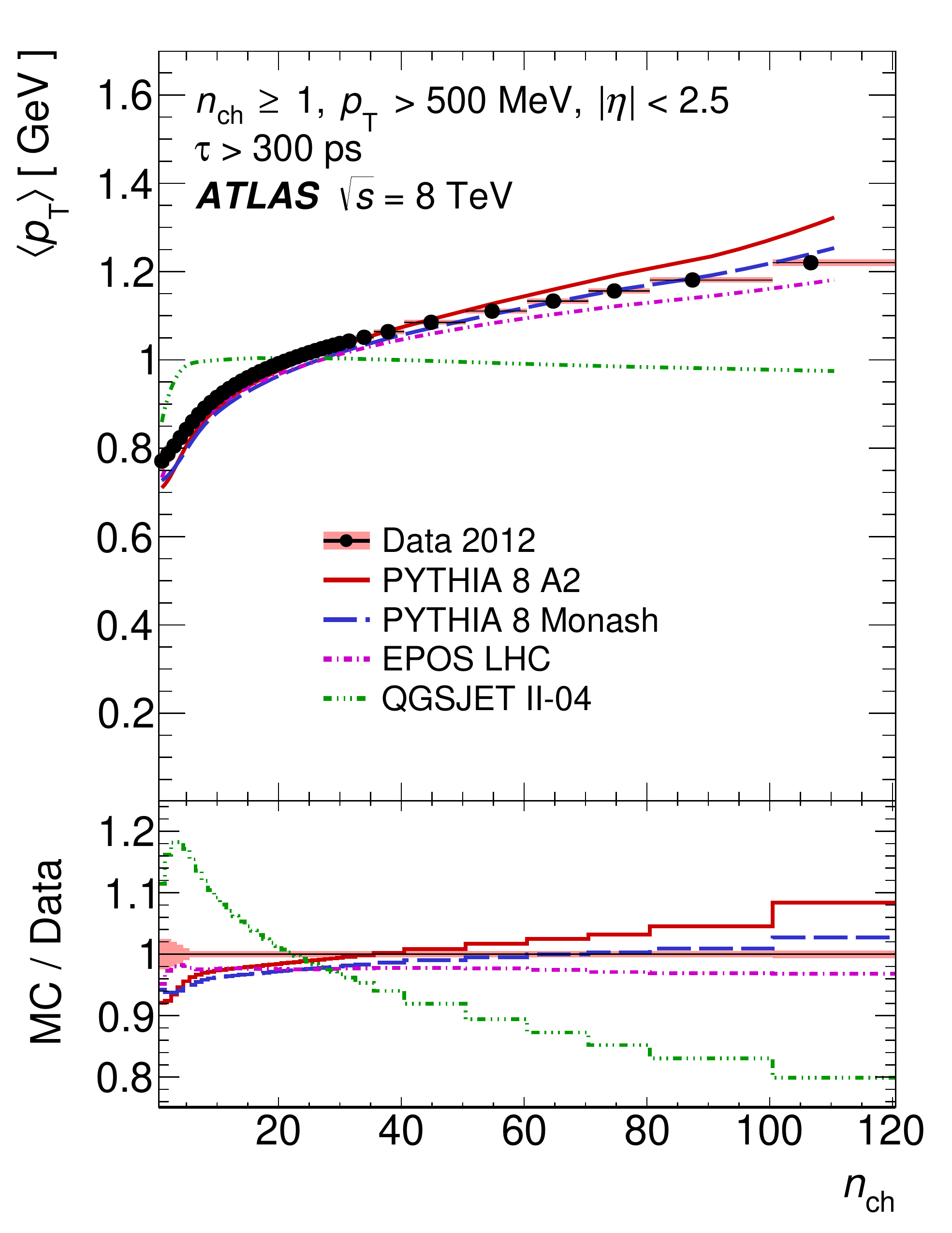}}
    \caption{Distributions of primary charged particles in events for which $\nch \geq 1$, $\pT > 500 \MeV$ and $|\eta| < 2.5$ as a function of (a) pseudorapidity, $\eta$, (b) transverse momentum, \pT, (c) multiplicity, \nch, and (d) average transverse momentum, \meanpT, versus multiplicity.
The data, represented by dots, are compared to various particle-level MC predictions, which are shown by curves.
The shaded areas around the data points represent the total statistical and systematic uncertainties added in quadrature.
\label{fig:final_500_1}}
\end{figure}

\begin{figure}[htbp]
\centering
    \subfloat[\label{fig:eta_500_6}]
{\includegraphics[width=0.48\textwidth]{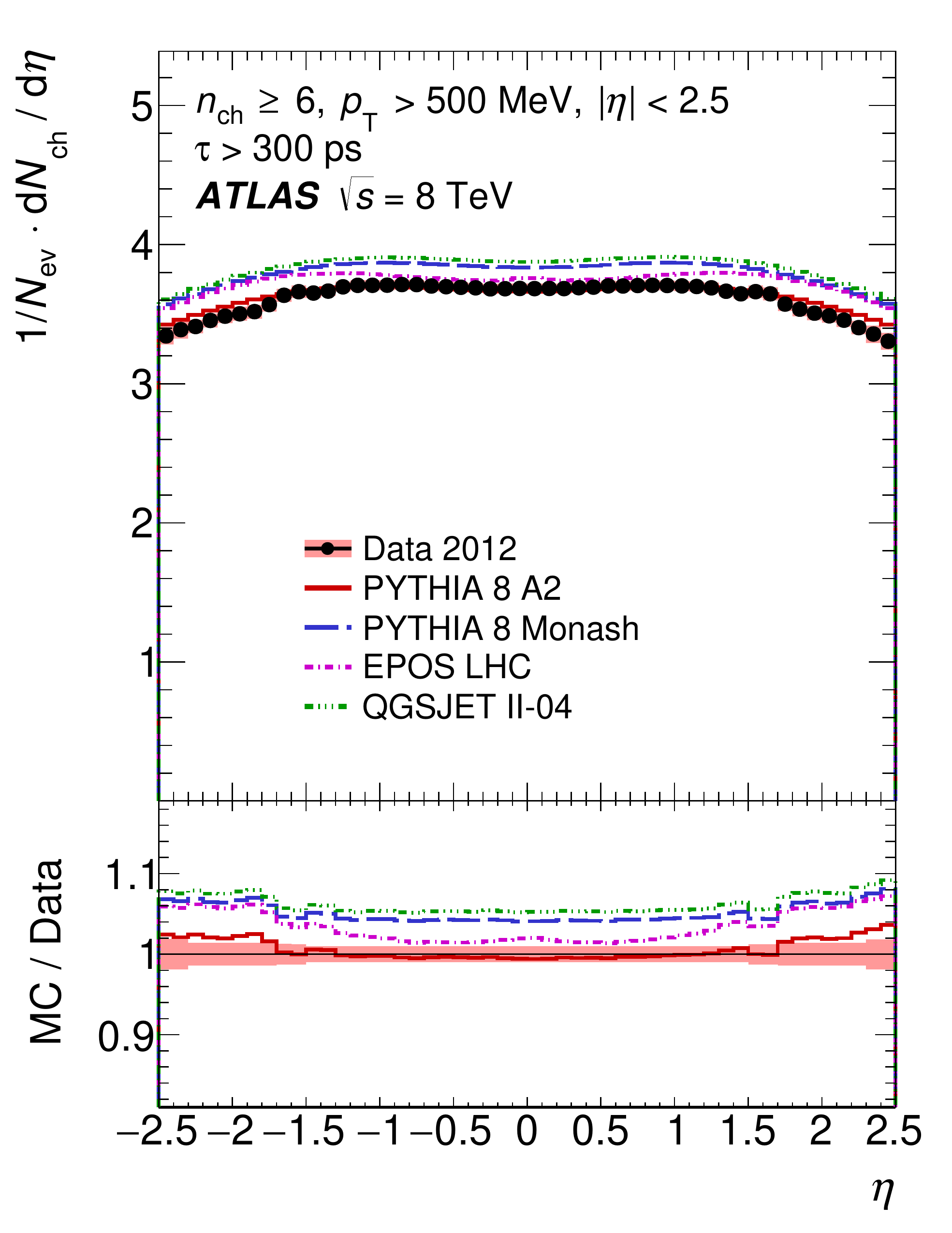}}
    \subfloat[\label{fig:pt_500_6}]
{\includegraphics[width=0.48\textwidth]{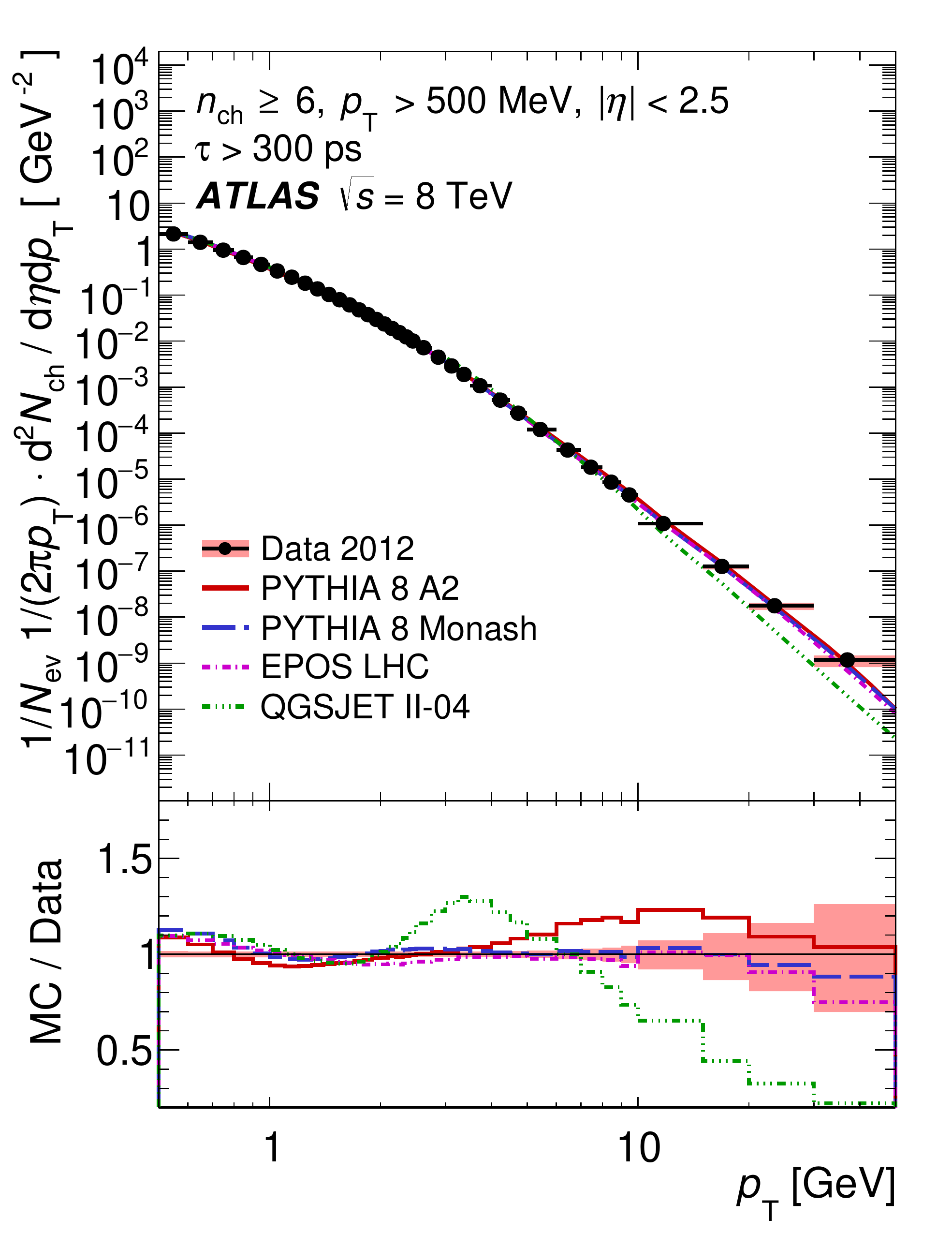}}\\
    \caption{Distributions of primary charged particles in events for which $\nch \geq 6$, $\pT > 500 \MeV$ and $|\eta| < 2.5$ as a function of (a) pseudorapidity, $\eta$, and (b) transverse momentum, \pT.
The data, represented by dots, are compared to various particle-level MC predictions, which are shown by curves.
The shaded areas around the data points represent the total statistical and systematic uncertainties added in quadrature.
\label{fig:final_500_6}}
\end{figure}

\begin{figure}[htbp]
\centering
    \subfloat[\label{fig:eta_500_20}]
{\includegraphics[width=0.48\textwidth]{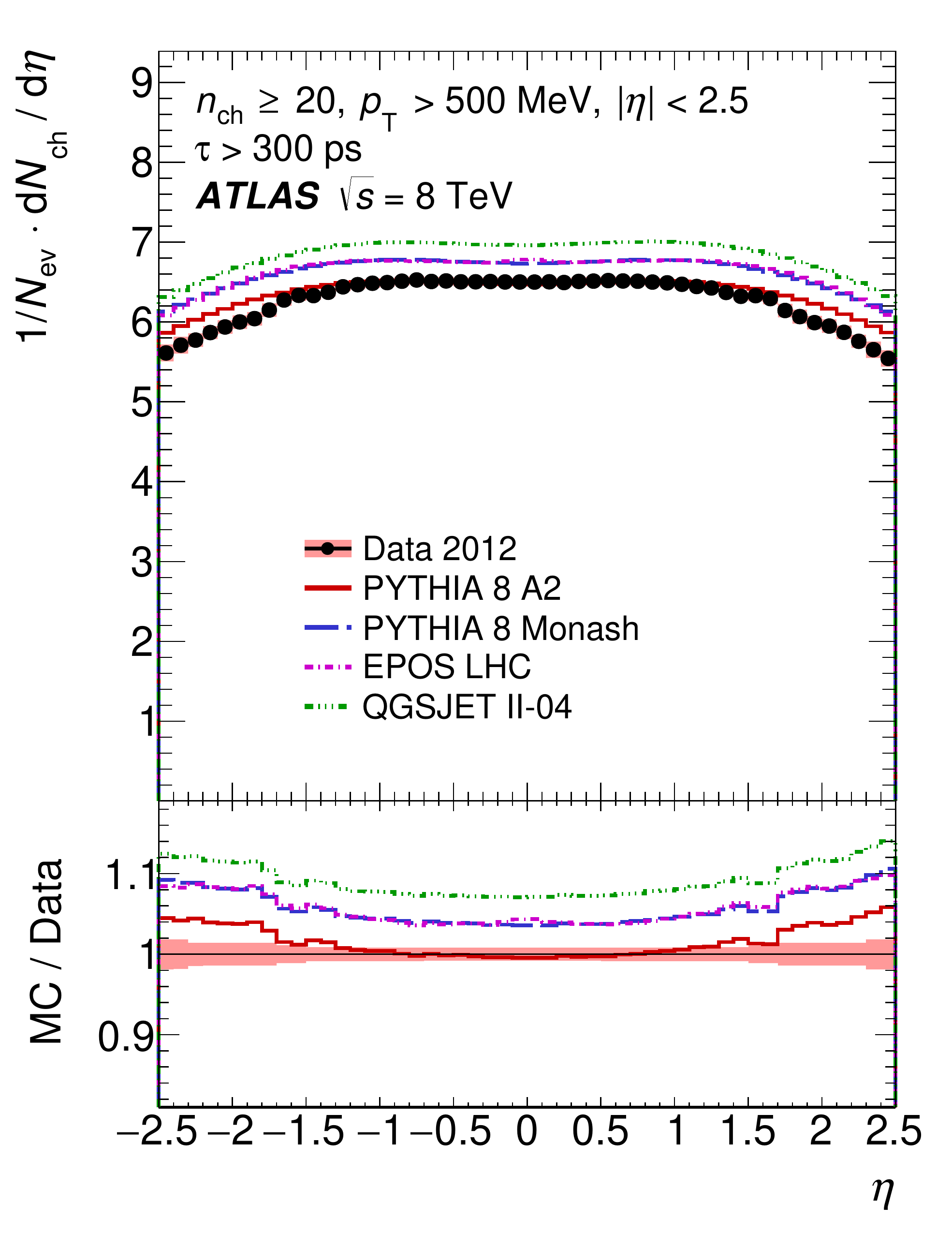}}
    \subfloat[\label{fig:pt_500_20}]
{\includegraphics[width=0.48\textwidth]{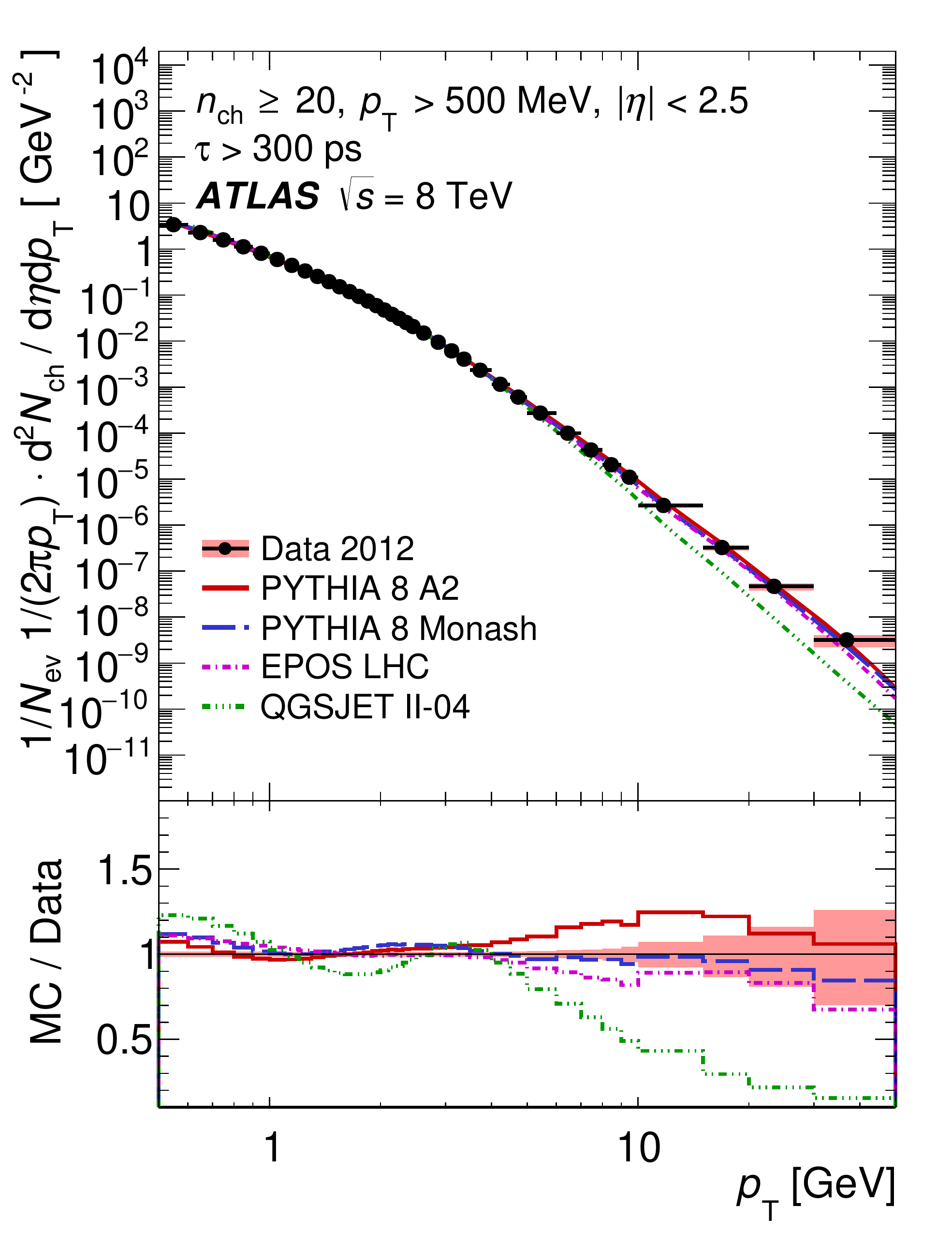}}\\
    \caption{Distributions of primary charged particles in events for which $\nch \geq 20$, $\pT > 500 \MeV$ and $|\eta| < 2.5$ as a function of (a) pseudorapidity, $\eta$, and (b) transverse momentum, \pT.
The data, represented by dots, are compared to various particle-level MC predictions, which are shown by curves.
The shaded areas around the data points represent the total statistical and systematic uncertainties added in quadrature.
\label{fig:final_500_6}}
\end{figure}

\begin{figure}[htbp]
\centering
    \subfloat[\label{fig:eta_500_50}]
{\includegraphics[width=0.48\textwidth]{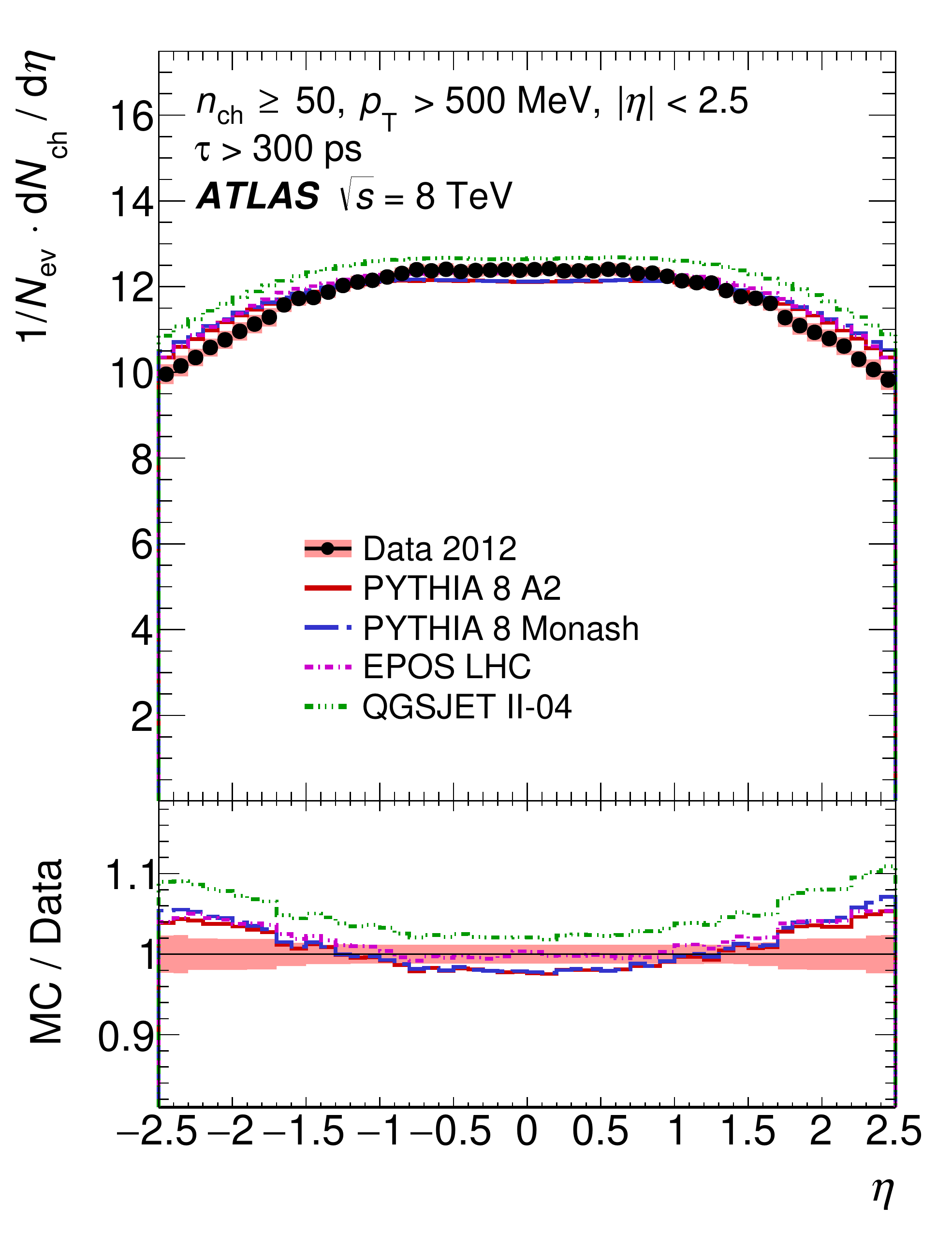}}
    \subfloat[\label{fig:pt_500_50}]
{\includegraphics[width=0.48\textwidth]{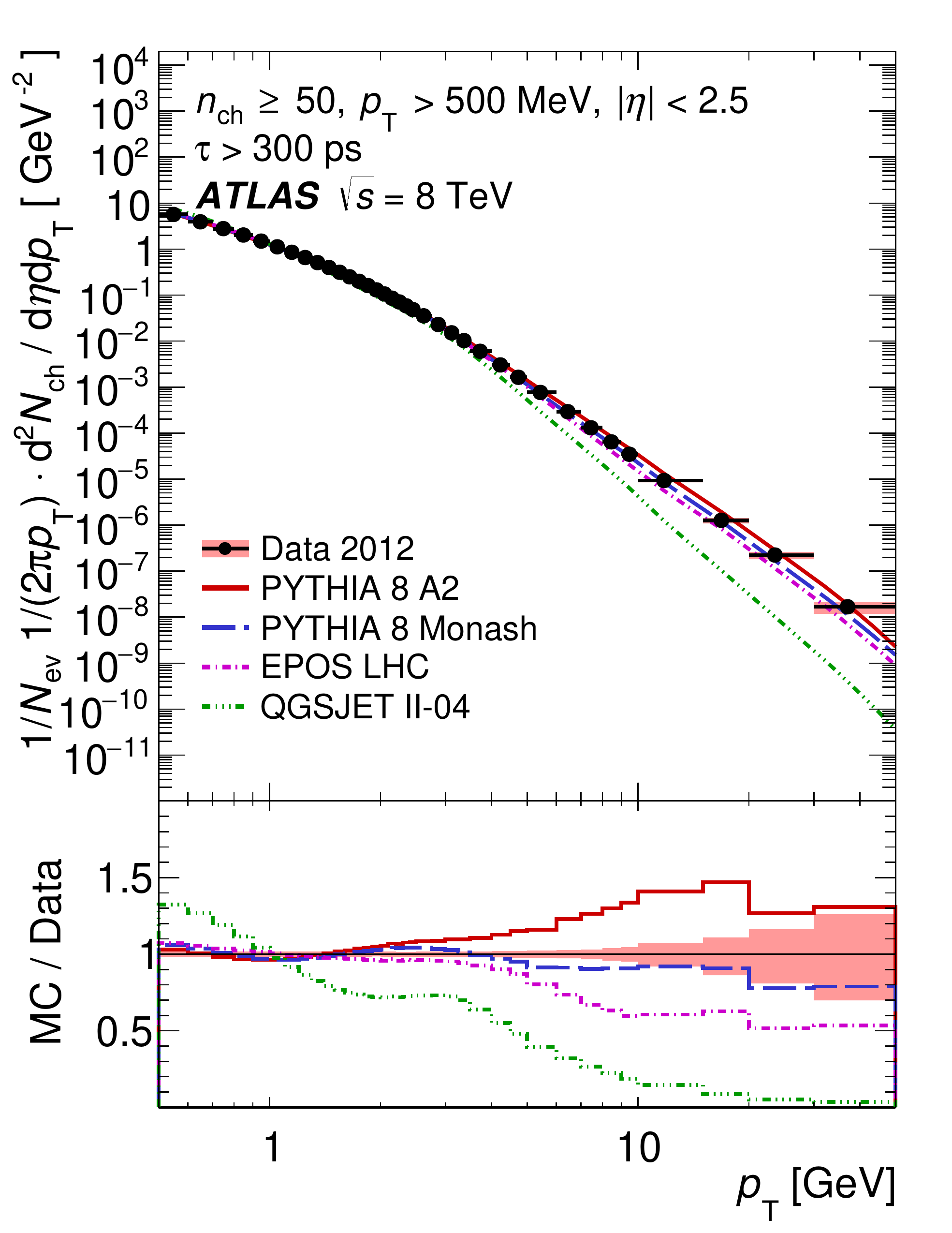}}\\
    \caption{Distributions of primary charged particles in events for which $\nch \geq 50$, $\pT > 500 \MeV$ and $|\eta| < 2.5$ as a function of (a) pseudorapidity, $\eta$, and (b) transverse momentum, \pT.
The data, represented by dots, are compared to various particle-level MC predictions, which are shown by curves.
The shaded areas around the data points represent the total statistical and systematic uncertainties added in quadrature.
\label{fig:final_500_50}}
\end{figure}

Distributions of primary-charged-particle pseudorapidity, \dNdeta, are given in
\Figs{\ref{fig:eta_100_2}, \ref{fig:eta_500_1}, \ref{fig:eta_500_6}, \ref{fig:eta_500_20}}{\ref{fig:eta_500_50}}
for all measured phase spaces. The distribution corresponding to the phase space $\nch \geq 2$ and $\pT > 100 \MeV$ rises as $|\eta|$ increases, peaking at $|\eta| \sim 2$ before falling. For the phase space $\nch \geq 1$ and $\pT > 500 \MeV$, the distribution is approximately constant for $|\eta|<2$ and falls at higher $|\eta|$. A similar shape is seen for the phase spaces requiring a higher multiplicity ($\nch \geq 6,20,50$) with the extent of the plateau becoming shorter as the multiplicity threshold is raised.  A small apparent structure in the distributions 
of the central values of the data points occurs at values of $|\eta| \sim 1.7$. This is due to systematic effects in the track
reconstruction efficiency which arises due to assumptions on the ID material composition, and is thus covered by
the total systematic uncertainty (see \Sect{\ref{sec:systematics}}).

The distribution corresponding to the phase space $\nch \geq 2$ and $\pT > 100 \MeV$ is well described by \EPOSLHC and \PYTHIAmon but is underestimated by \PYTHIAatwo and \QGSJET.\footnote{%
The MC models used here were mostly tuned to data in the $\pT > 500 \MeV$ phase space (up to $\sqrt{s} = 7 \TeV$) and can therefore not necessarily be expected to describe the distributions at $\pT > 100 \MeV$ as well as at $\pT > 500 \MeV$.} %
For the phase space $\nch \geq 1$ and $\pT > 500 \MeV$, \EPOSLHC overestimates the distribution at values of $|\eta| > 1.7$ and describes the data well for the rest of the pseudorapidity range.  The data are overestimated by the \QGSJET and \PYTHIAmon calculations and underestimated by the \PYTHIAatwo prediction. All models overestimate the overall yield for the phase spaces $\nch \geq 6,20$ although \PYTHIAatwo describes the plateau in the central region well. For the largest multiplicity threshold ($\nch \geq 50$) all of the models overestimate the data at $|\eta| > 1.7$ but provide a better description in the central region.

\Figs{\ref{fig:pt_100_2}, \ref{fig:pt_500_1}, \ref{fig:pt_500_6}, \ref{fig:pt_500_20}}{\ref{fig:pt_500_50}}
show distributions of primary-charged-particle transverse momentum, \dNdpT, for various phase spaces.
No model is fully consistent with the distributions, although above $1 \GeV$ the \PYTHIAmon predictions agree well with the data. This is also the only model which gives a fair description of the data corresponding to the highest multiplicity threshold with $\nch \geq 50$ and $\pT > 500 \MeV$, where all other models show large deviations as \pT increases.
The \EPOSLHC predictions give the best description of the data corresponding to the phase space $\nch \geq 2$ and $\pT > 100 \MeV$, particularly at transverse momenta below $1 \GeV$, while the other models underestimate the data at the lowest \pT values. \EPOSLHC provides fair predictions for the phase spaces $\nch \geq 1, 6$ and $\pT > 500 \MeV$, but for the higher multiplicity thresholds ($\nch \geq 20$ and 50) deviations from the data are seen at high transverse momenta.
\PYTHIAatwo gives fair descriptions of the data below $6 \GeV$, yet shows deviations of up to \SI{30}{\percent} around $\pT \sim 10 \GeV$.
In all measured phase spaces, the \QGSJET approach shows large disagreements with the data as \pT increases.

In \Figs{\ref{fig:nch_100_2}}{\ref{fig:nch_500_1}} distributions of primary-charged-particle multiplicity, \dnchnorm, are shown for minimum transverse momentum thresholds of $100 \MeV$ and $500 \MeV$, respectively.
For the lower threshold, the distribution rises until values of $\nch \sim 9$ before falling steeply. For the higher threshold the distribution peaks at $\nch \sim 2$. None of the models are consistent with the data although the \EPOSLHC model provides a fair description. The two \PYTHIA 8 calculations predict distribution peaks which are at higher \nch than those observed and underestimate the event yield at low and high multiplicity. The \QGSJET tune overestimates the data at low and high \nch values and underestimates the data for intermediate \nch values.

The distribution of the average transverse momentum of primary charged particles, \meanpT, versus the primary-charged-particle multiplicity, \nch, is given in
\Figs{\ref{fig:avgpt_100_2}}{\ref{fig:avgpt_500_1}} for transverse momentum thresholds of $100 \MeV$ and $500 \MeV$, respectively.
The average \pT rises with multiplicity although the rise becomes progressively less steep as the multiplicity increases. This is expected due to colour coherence effects in dense parton environments, which are modelled by a colour reconnection mechanism in \PYTHIA 8 or by the hydrodynamical evolution model used in \EPOS. It is assumed that numerous MPI dominate the high-multiplicity events, and that colour coherence effects thereby lead to fewer additional charged particles produced with every additional MPI, which share a higher average \pT. The \EPOSLHC and \PYTHIA 8 models provide a fair description of the data.
The \QGSJET model fails to predict the mean transverse momentum over the entire multiplicity range, as it does not simulate colour coherence effects and therefore shows very little dependence on the multiplicity.


\begin{figure}[htbp]
\centering
    \subfloat[]{\includegraphics[width=0.49\textwidth]{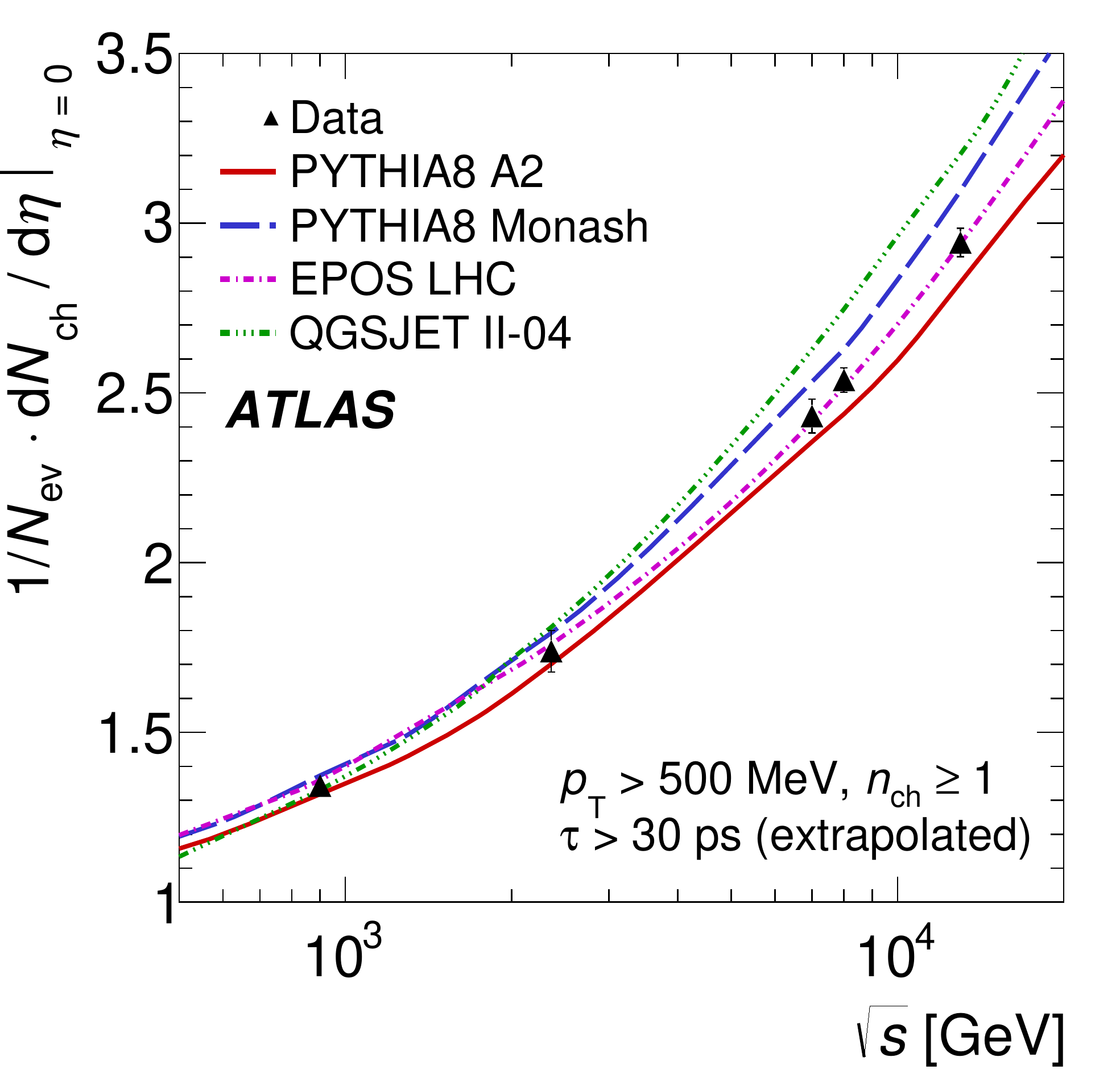}}
    \subfloat[]{\includegraphics[width=0.49\textwidth]{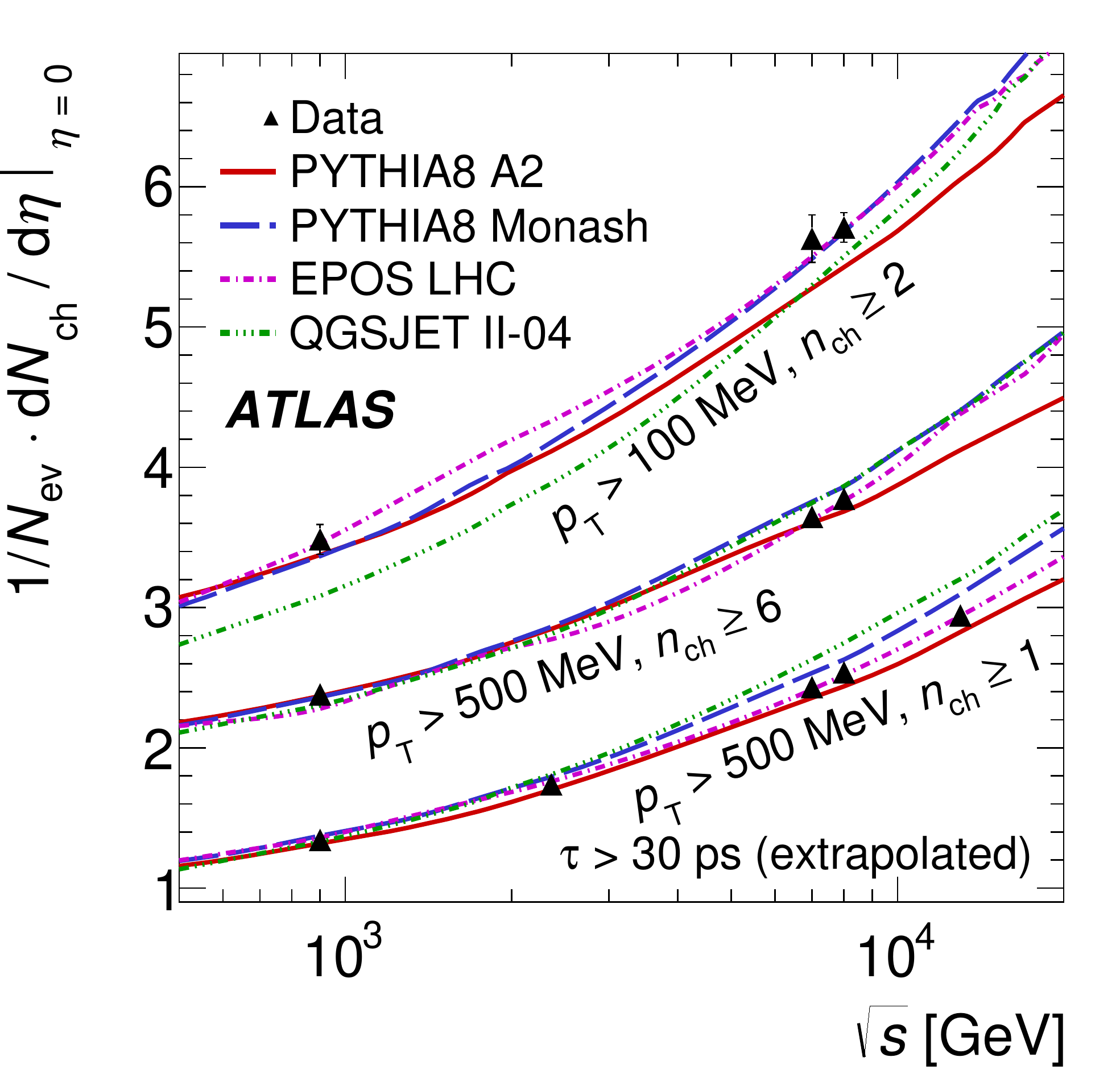}}
    \caption{The average primary-charged-particle multiplicity per unit of pseudorapidity at $\eta = 0$ as a function of the centre-of-mass energy. Results are shown for the phase spaces (a) ($\pT > 500 \MeV$, $\nch \geq 1$) and (b) ($\pT > 500 \MeV$, $\nch \geq 1$), ($\pT > 500 \MeV$, $\nch \geq 6$), and ($\pT > 100 \MeV$, $\nch \geq 2$). The data are compared to various particle-level MC predictions.
The results at $\sqrt{s} = 8$ and $13 \TeV$ are extrapolated to include strange baryons.
The vertical error bars on the data represent the total uncertainty.
\label{fig:sqrts}}
\end{figure}

\begin{table}[ht!]
  \centering
    \begin{tabular}{
    c
    c
    c
    c
    }
\toprule
\multicolumn{2}{c}{\bf Phase Space} 	& \multicolumn{2}{c}{\bf \dNdeta at $\eta = 0$}  \\
{$\nch \geq$}  & {$\pT >$} 	& {$\tau > \SI{300}{\ps}$ (fiducial)}	& {$\tau > \SI{30}{\ps}$ (extrapolated)} \\
\midrule
2 			&  100 \MeV 		&  {5.64~~ $\pm$ 0.10~~} 	&  5.71~~ $\pm$ 0.11  \\
1 			&  500 \MeV		&  {2.477 $\pm$ 0.031} 		&  2.54~~ $\pm$ 0.04  \\
6 			&  500 \MeV		&  {3.68~~ $\pm$ 0.04~~} 	&  3.78~~ $\pm$ 0.05  \\
20 			&  500 \MeV		&  {6.50~~ $\pm$ 0.05~~} 	&  6.66~~ $\pm$ 0.07  \\
50 			&  500 \MeV		&  {12.40~~ $\pm$ 0.15~~~~} 	&  12.71~~ $\pm$ 0.18~~  \\
\bottomrule
    \end{tabular}
    \caption{
Central primary-charged-particle density \dNdeta at $\eta = 0$ for five different phase spaces.
The results are given for the fiducial definition {$\tau > \SI{300}{\ps}$}, as well as for the previously used fiducial definition {$\tau > \SI{30}{\ps}$} using an extrapolation factor of
$1.012 \pm 0.004$ (for $\pT > 100 \MeV$) or $1.025 \pm 0.008$ (for $\pT > 500 \MeV$),
which accounts for the fraction of charged strange baryons predicted by \EPOSLHC simulation.
    \label{tab:dndeta_0_all}}
\end{table}

The evolution of the primary-charged-particle multiplicity per unit pseudorapidity at $\eta=0$ is shown in \Fig{\ref{fig:sqrts}}.
It is computed by averaging over $|\eta|<0.2$ in the \dNdeta distribution.
In order to make consistent comparisons with previous measurements,
these figures are corrected to the earlier $\tau > \SI{30}{\ps}$ definition of stable particles
(to include the fraction of short-lived particles which have been excluded from this study),
using a factor
$1.012 \pm 0.004$ in the $\pT > 100 \MeV$ phase space and
$1.025 \pm 0.008$ in the $\pT > 500 \MeV$ phase spaces, derived from predictions of the \EPOSLHC
tune with uncertainties following comparisons of the predictions of different MC models.
Results are shown for the phase spaces ($\pT > 500 \MeV$, \ $\nch \geq 1$), ($\pT > 500 \MeV$, \ $\nch \geq 6$), and ($\pT > 100 \MeV$, \ $\nch \geq 2$) along with available results from other ATLAS measurements at $\sqrt{s} = 0.9, 2.36, 7$ and $13 \TeV$~\cite{Aad:2016mok,Aad:2010ac,Aad:2010rd}.
It can be seen that the total uncertainty in the measurement at $\sqrt{s} = 8 \TeV$ is about \SIrange[range-phrase = --]{30}{40}{\percent} less than for the study with $7 \TeV$ data~\cite{Aad:2010ac}.
This was achieved due to our improved knowledge of the ID material distribution~\cite{Lukas:1956718}, which reduced the dominant source of systematic uncertainty by more than \SI{50}{\percent} with respect to the previous $7 \TeV$ measurement.
Predictions of various QCD-based models are also shown. The best description of the data is given by \EPOSLHC. The predictions of the \PYTHIA 8 tunes provide a fair description of the shape of the multiplicity dependence with centre-of-mass energy. As in the case of the other presented distributions, calculations of \QGSJET give the worst description.

A full summary of central primary-charged-particle densities at $\eta=0$ in all measured phase spaces is given in \Tab{\ref{tab:dndeta_0_all}}, showing results obtained with the new as well as the previous fiducial definition.


\FloatBarrier

\section{Conclusion}
\label{sec:conclusion}

Measurements were made of distributions of primary charged particles produced in minimum-bias $pp$ collisions at $\sqrt{s} = 8 \TeV$ with the ATLAS detector at the LHC. The results are based on a dataset corresponding to an integrated luminosity of \SI{160}{\per\micro\barn}.
Distributions of primary-charged-particle multiplicities as well as pseudorapidity and transverse momentum spectra are shown.
With the fiducial definition of primary charged particles that was used in this study ($\tau > \SI{300}{\ps}$), the central primary-charged-particle multiplicity at $\eta = 0$ per event and unit of pseudorapidity was measured to be $5.64 \pm 0.10$ in events containing $\nch \geq 2$ primary charged particles with transverse momentum $\pT > 100 \MeV$, and
$2.477 \pm 0.031$ in events with $\nch \geq 1$ and $\pT > 500 \MeV$.
Using an extrapolation factor for short-lived charged particles with a lifetime between $30 < \tau < \SI{300}{\ps}$, the central primary-charged-particle multiplicity was measured to be $5.71 \pm 0.11$ and $2.54 \pm 0.04$, respectively.
The precision of these results is \SIrange[range-phrase = --]{30}{40}{\percent} better than for the previous highest precision ATLAS measurements at 0.9 and $7 \TeV$.
Compared with earlier studies, this paper also presents ATLAS measurements of final states at high multiplicities of $\nch \geq 20$ and $\nch \geq 50$.
Predictions of various Monte Carlo models were compared with the data, and it was found that the best description is given by the \EPOSLHC tune, followed by the \PYTHIAatwo and \textsc{Monash} tunes.
The measurements presented here are expected to provide valuable constraints for the tuning and further understanding of soft QCD physics models.

\section*{Acknowledgements}


We thank CERN for the very successful operation of the LHC, as well as the
support staff from our institutions without whom ATLAS could not be
operated efficiently.

We acknowledge the support of ANPCyT, Argentina; YerPhI, Armenia; ARC, Australia; BMWFW and FWF, Austria; ANAS, Azerbaijan; SSTC, Belarus; CNPq and FAPESP, Brazil; NSERC, NRC and CFI, Canada; CERN; CONICYT, Chile; CAS, MOST and NSFC, China; COLCIENCIAS, Colombia; MSMT CR, MPO CR and VSC CR, Czech Republic; DNRF and DNSRC, Denmark; IN2P3-CNRS, CEA-DSM/IRFU, France; GNSF, Georgia; BMBF, HGF, and MPG, Germany; GSRT, Greece; RGC, Hong Kong SAR, China; ISF, I-CORE and Benoziyo Center, Israel; INFN, Italy; MEXT and JSPS, Japan; CNRST, Morocco; FOM and NWO, Netherlands; RCN, Norway; MNiSW and NCN, Poland; FCT, Portugal; MNE/IFA, Romania; MES of Russia and NRC KI, Russian Federation; JINR; MESTD, Serbia; MSSR, Slovakia; ARRS and MIZ\v{S}, Slovenia; DST/NRF, South Africa; MINECO, Spain; SRC and Wallenberg Foundation, Sweden; SERI, SNSF and Cantons of Bern and Geneva, Switzerland; MOST, Taiwan; TAEK, Turkey; STFC, United Kingdom; DOE and NSF, United States of America. In addition, individual groups and members have received support from BCKDF, the Canada Council, CANARIE, CRC, Compute Canada, FQRNT, and the Ontario Innovation Trust, Canada; EPLANET, ERC, FP7, Horizon 2020 and Marie Sk{\l}odowska-Curie Actions, European Union; Investissements d'Avenir Labex and Idex, ANR, R{\'e}gion Auvergne and Fondation Partager le Savoir, France; DFG and AvH Foundation, Germany; Herakleitos, Thales and Aristeia programmes co-financed by EU-ESF and the Greek NSRF; BSF, GIF and Minerva, Israel; BRF, Norway; Generalitat de Catalunya, Generalitat Valenciana, Spain; the Royal Society and Leverhulme Trust, United Kingdom.

The crucial computing support from all WLCG partners is acknowledged
gratefully, in particular from CERN and the ATLAS Tier-1 facilities at
TRIUMF (Canada), NDGF (Denmark, Norway, Sweden), CC-IN2P3 (France),
KIT/GridKA (Germany), INFN-CNAF (Italy), NL-T1 (Netherlands), PIC (Spain),
ASGC (Taiwan), RAL (UK) and BNL (USA) and in the Tier-2 facilities
worldwide.

%


\clearpage
\appendix
\part*{Appendix}
\addcontentsline{toc}{part}{Appendix}

\section{Average number of measurements per track}
\label{sec:hitsontrack}

\begin{figure}[htbp]
\centering
    \subfloat[\label{fig:auxhits_blayer}]{\includegraphics[width=0.48\textwidth]{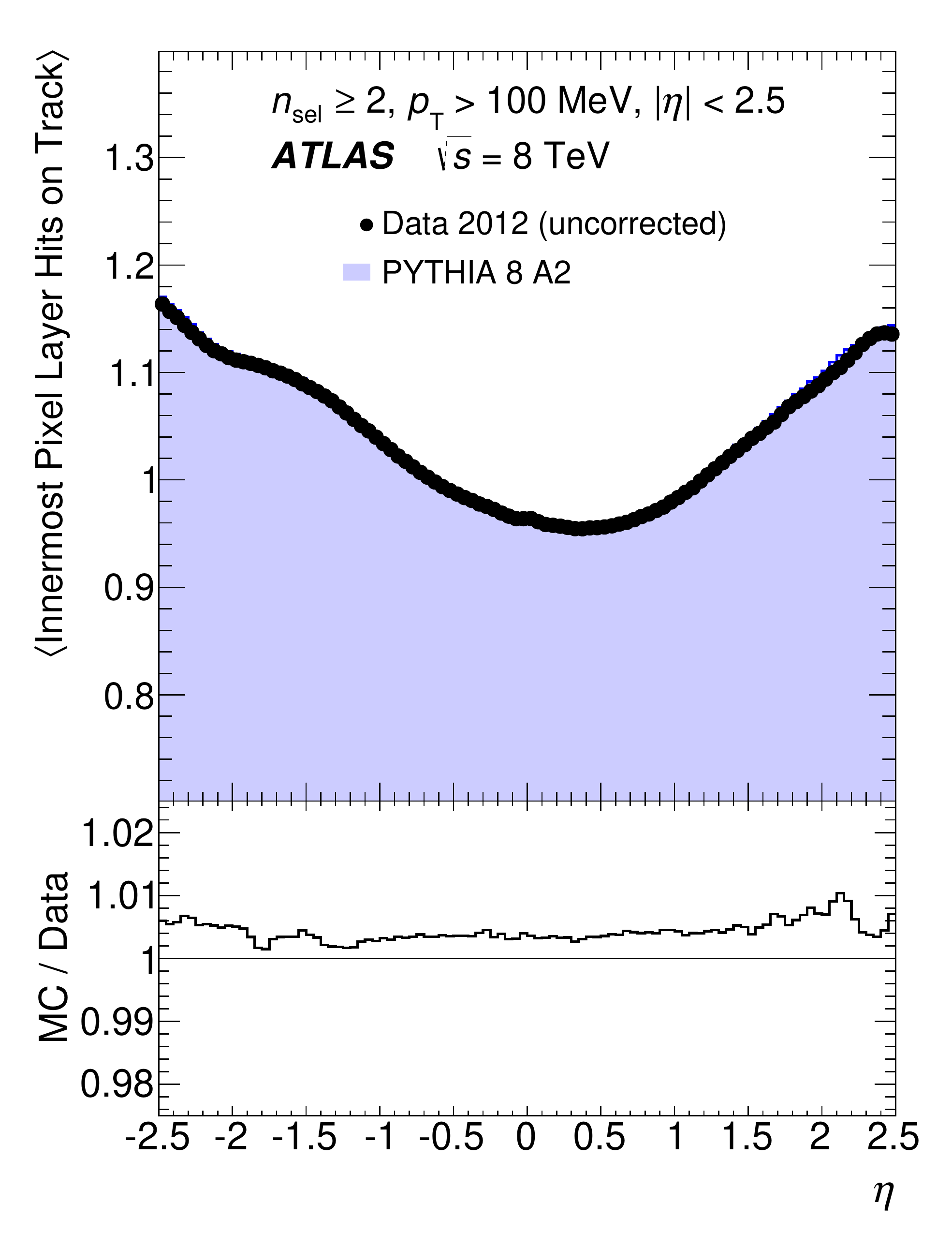}}
    \subfloat[\label{fig:auxhits_pixel}]{\includegraphics[width=0.48\textwidth]{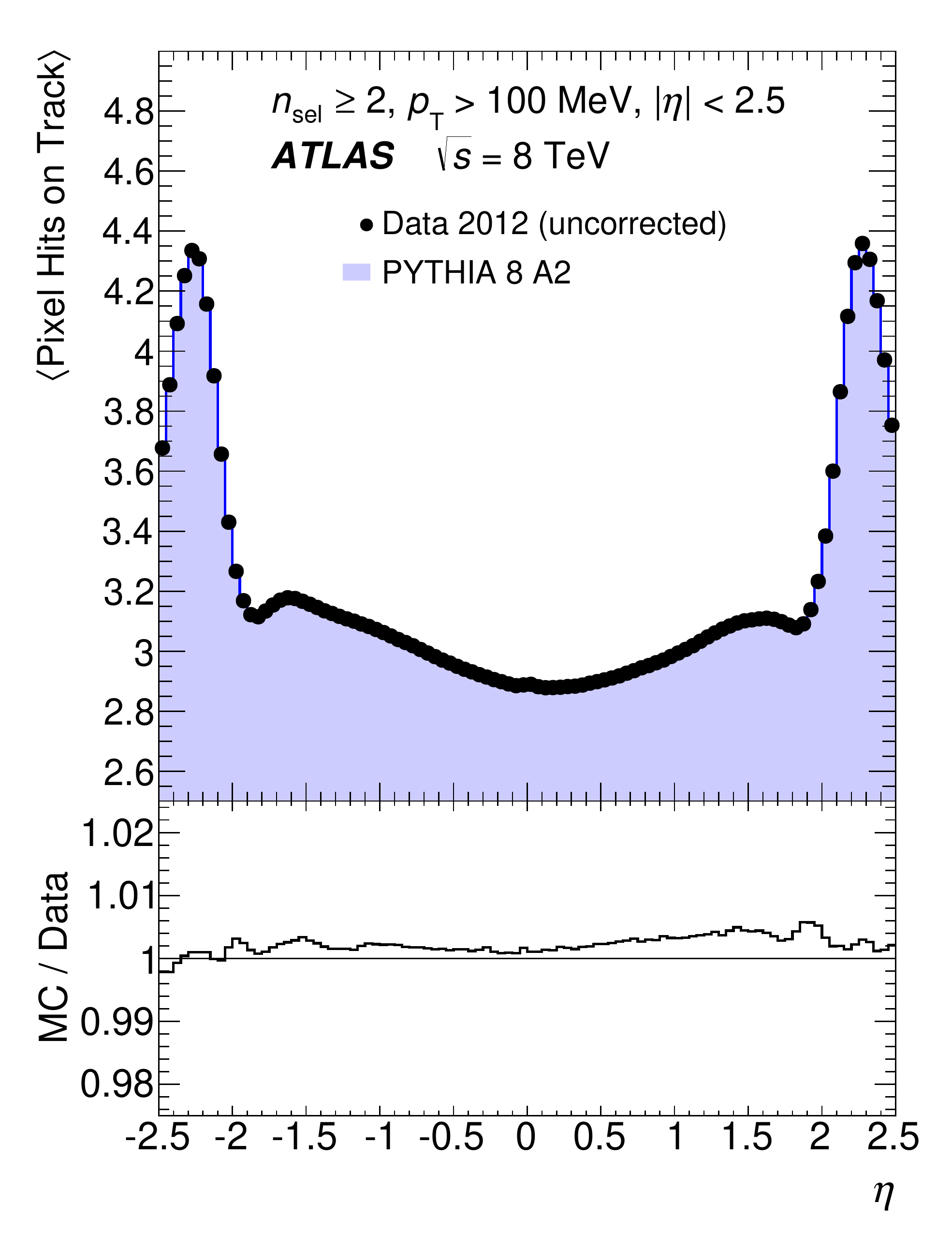}} \\
    \subfloat[\label{fig:auxhits_sct}]{\includegraphics[width=0.48\textwidth]{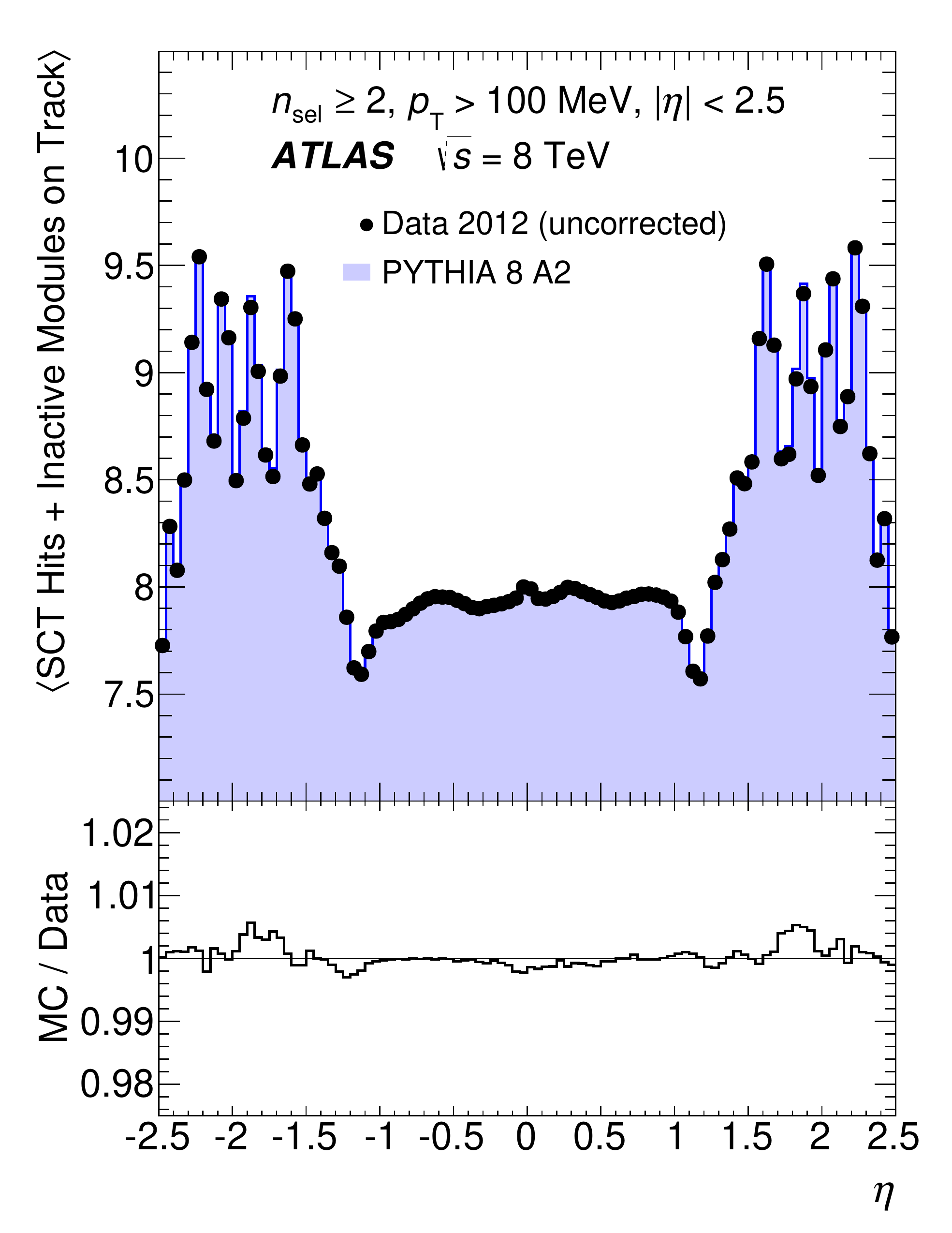}}
    \caption{Data and MC simulation distributions of the average number of hits per reconstructed track as a function of pseudorapidity, $\eta$, in (a) the innermost layer of the Pixel detector, (b) the whole Pixel detector, and (c) the SCT detector (adding the average number of hits and inactive modules per track). 
The MC simulation distributions, made with \PYTHIA 8 using the {\sc A2} tune, have been reweighted to match the reconstructed \pT spectrum in data.
    \label{fig:auxhits}}
\end{figure}

The distributions of the average number of hits per reconstructed track in data and MC simulation are shown in \Fig{\ref{fig:auxhits}} for several detector components, using events selected by the ($\nsel \geq 2$, $\pT > 100 \MeV$) requirement. 
The distributions are shown as a function of the pseudorapidity of the reconstructed tracks. 
The MC simulation distributions, made with \PYTHIA 8 using the {\sc A2} tune, have been reweighted to match the reconstructed \pT spectrum in data.

The level of agreement between data and MC simulation is found to be within $\pm \SI{1}{\percent}$ for the average number of measurements per track in the innermost layer of the Pixel detector (\Fig{\ref{fig:auxhits_blayer}}), and remains within $\pm \SI{0.6}{\percent}$ for the average number of measurements per track in the whole Pixel detector (\Fig{\ref{fig:auxhits_pixel}}) as well as the SCT (\Fig{\ref{fig:auxhits_sct}}). 
For the SCT, the sum of average measurements per track and inactive modules per track is shown. This is done in order to account for a mis-modelling of inactive modules in the MC simulation, which was shown to have negligible impact on other results presented in this paper.

\clearpage

\section{Distributions of impact parameters}
\label{sec:app_ip}

\begin{figure}[htbp]
\centering
    \subfloat[]{\includegraphics[width=0.48\textwidth]{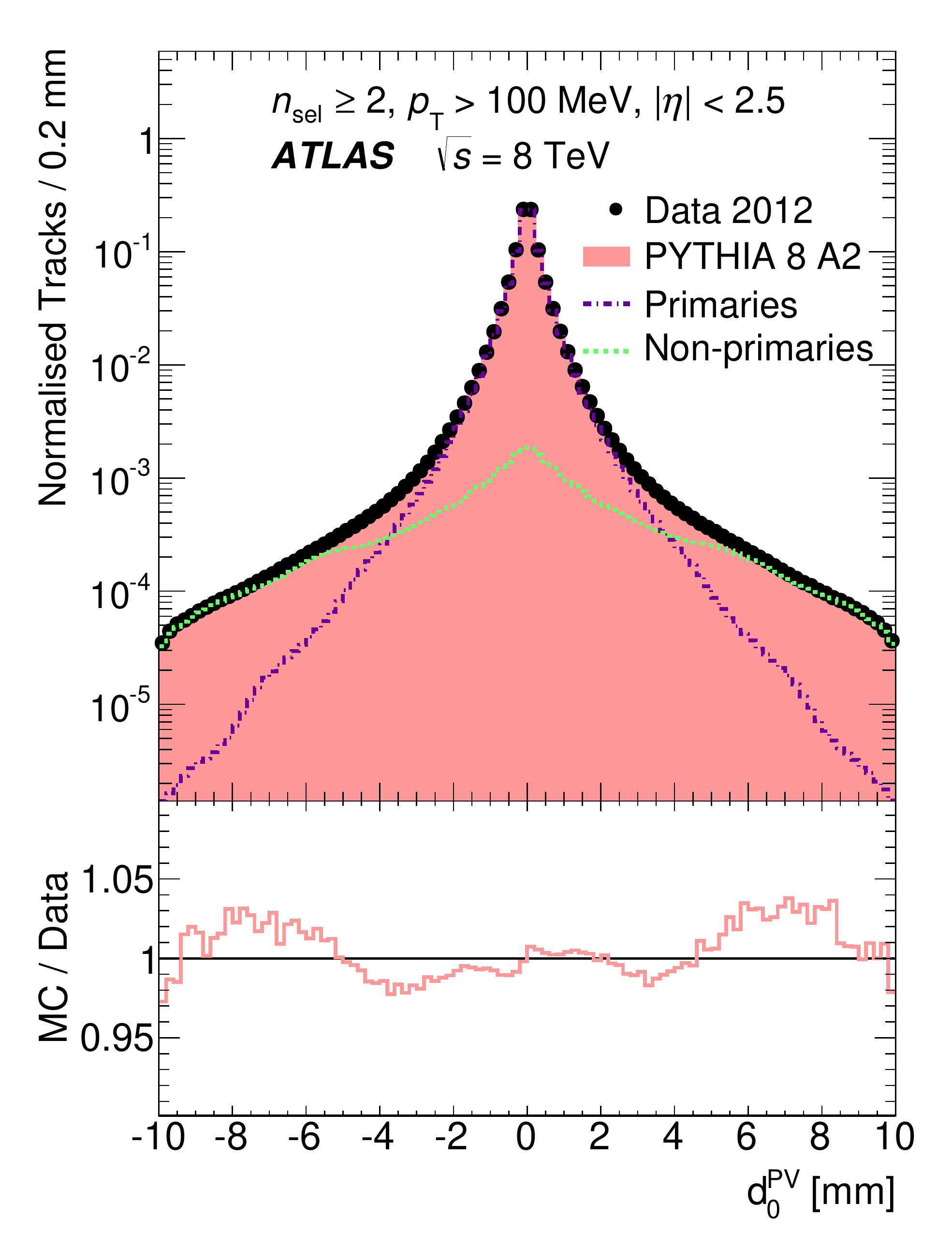}}
    \subfloat[]{\includegraphics[width=0.48\textwidth]{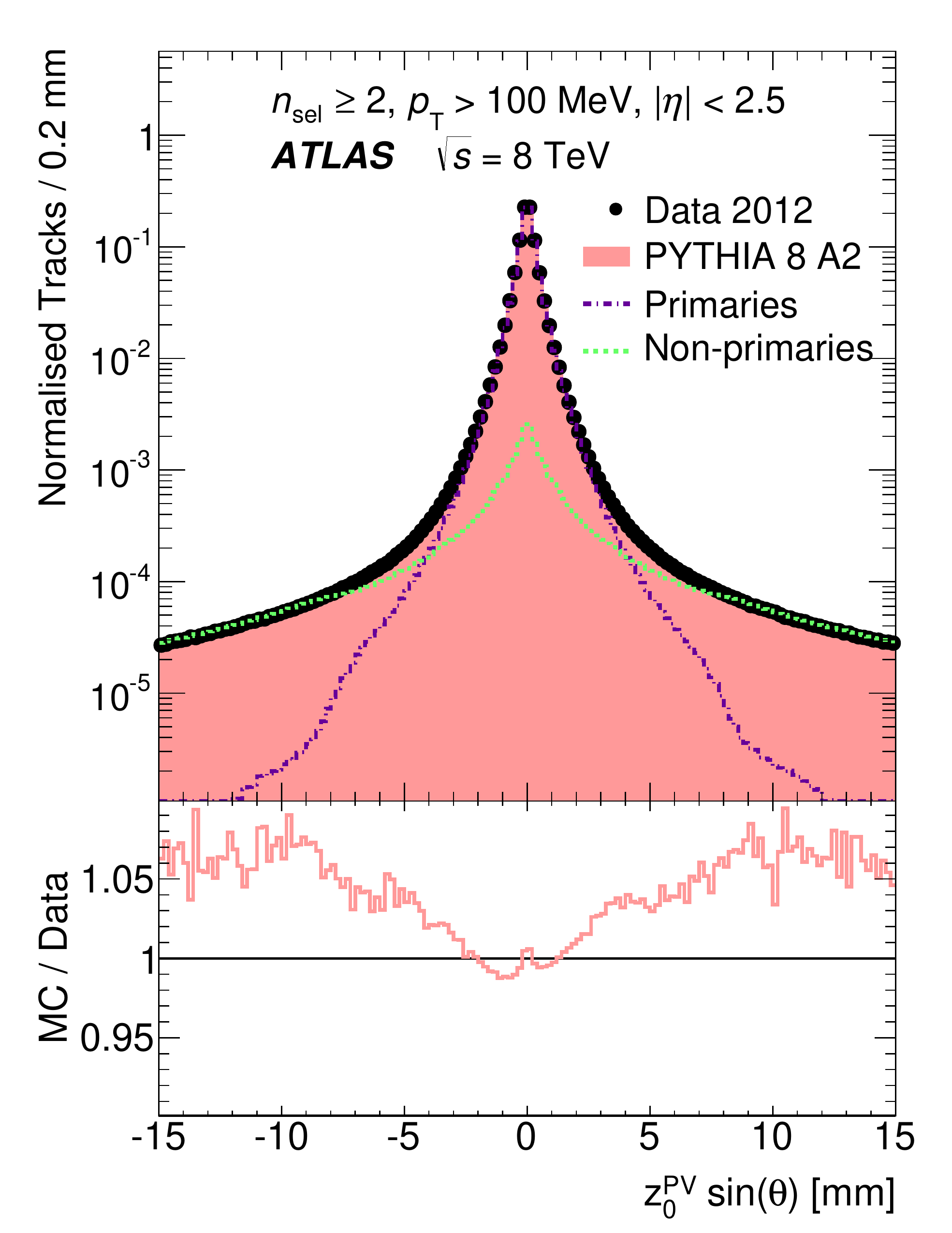}}
    \caption{Normalised distributions of the (a) transverse and (b) longitudinal impact parameters of reconstructed tracks in data and MC simulation with respect to the reconstructed primary vertex, 
using events selected by the ($\nsel \geq 2$, $\pT > 100 \MeV$) requirement. 
The MC simulation distributions, made with \PYTHIA 8 using the {\sc A2} tune, have been reweighted to match the reconstructed \pT spectrum as well as the fractions of reconstructed non-primary tracks in data.
    \label{fig:auxd0z0}}
\end{figure}

The normalised distributions of transverse and longitudinal impact parameters of reconstructed tracks in data and MC simulation with respect to the reconstructed primary vertex, using events selected by the ($\nsel \geq 2$, $\pT > 100 \MeV$) requirement, are shown in \Fig{\ref{fig:auxd0z0}}. 
The fractions of tracks originating from primary and secondary particles in the MC simulation (made with \PYTHIA 8 using the {\sc A2} tune), which have been reweighted to match the reconstructed \pT spectrum as well as the fractions of reconstructed non-primary tracks in data, are also shown. 

The level of agreement between data and MC simulation is found to be within $\pm \SI{1.5}{\percent}$ in the signal region of selected tracks used in the analysis, where the impact parameters are within $\pm \SI{1.5}{\mm}$ of the reconstructed primary vertex. In the tail regions, the level of agreement remains within $\pm \SI{4}{\percent}$ for the transverse impact parameter and within $\pm \SI{9}{\percent}$ for the longitudinal impact parameter. 

\clearpage

\section{Additional trigger efficiency plots}
\label{sec:trigeffplots}

\begin{figure}[htbp]
\centering
    \subfloat[\label{fig:trigeff_100_2}]{\includegraphics[width=0.48\textwidth]{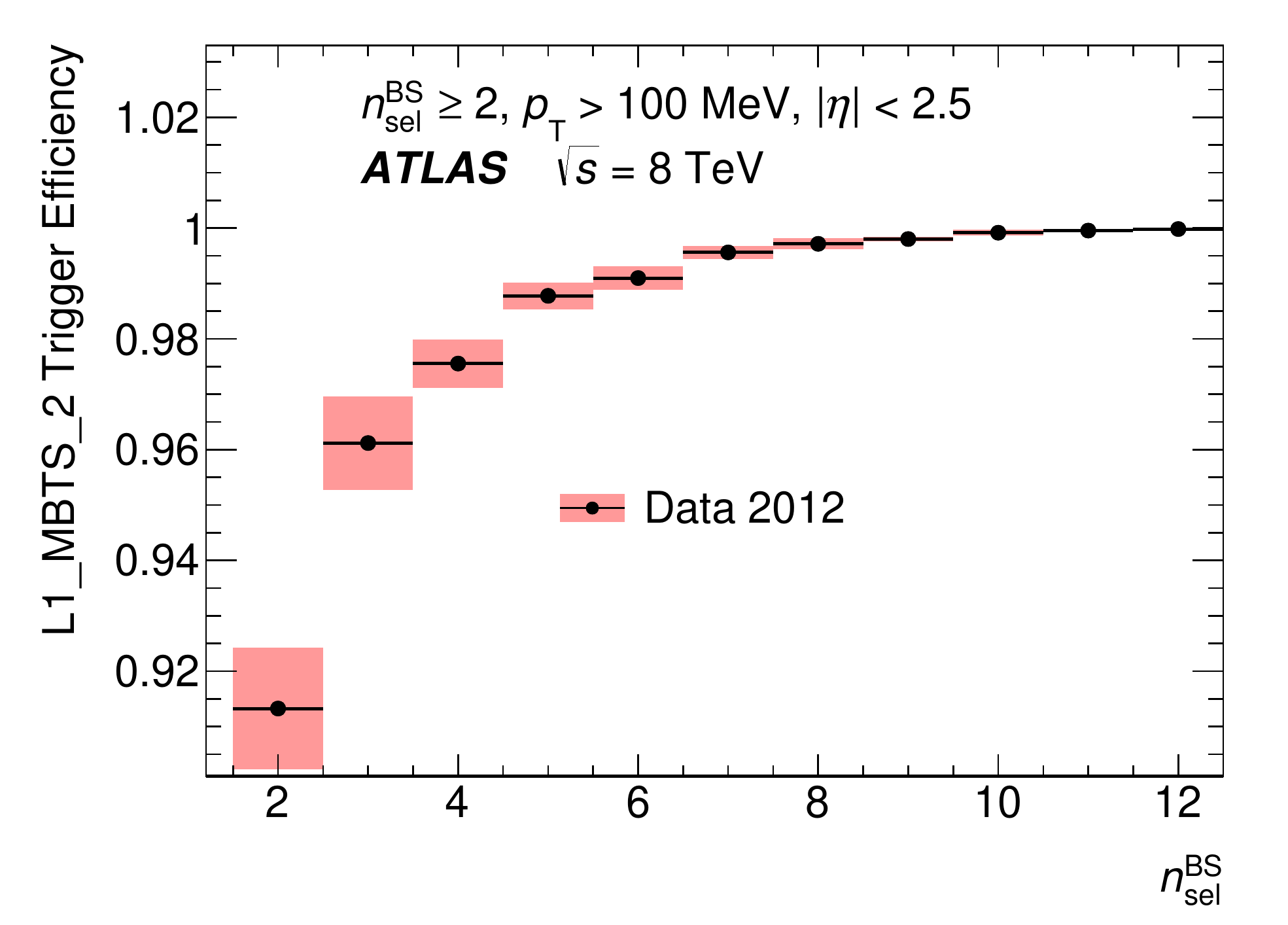}} \\
    \subfloat[\label{fig:trigeff_500_1}]{\includegraphics[width=0.48\textwidth]{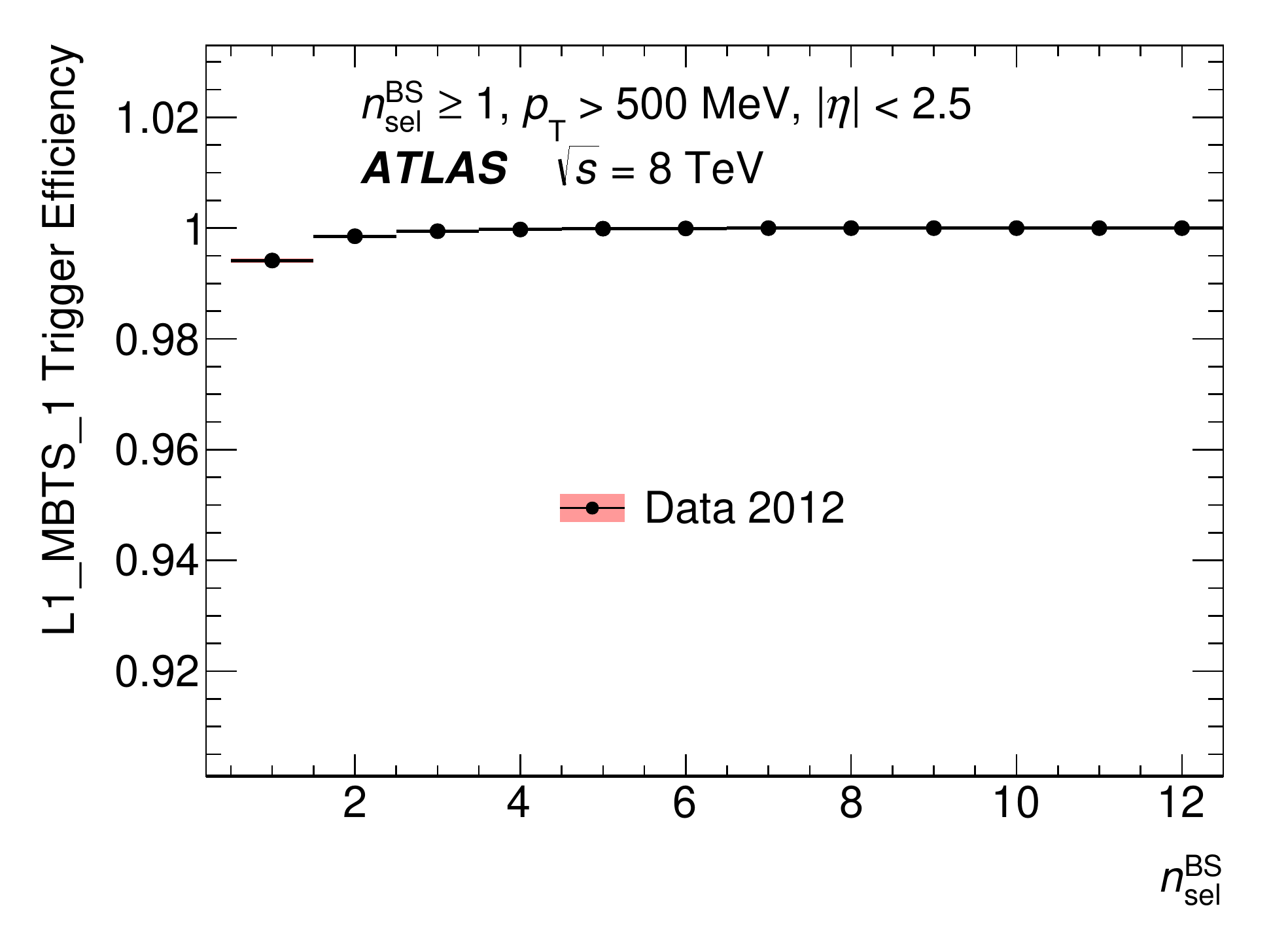}}
    \subfloat[\label{fig:trigeff_500_2}]{\includegraphics[width=0.48\textwidth]{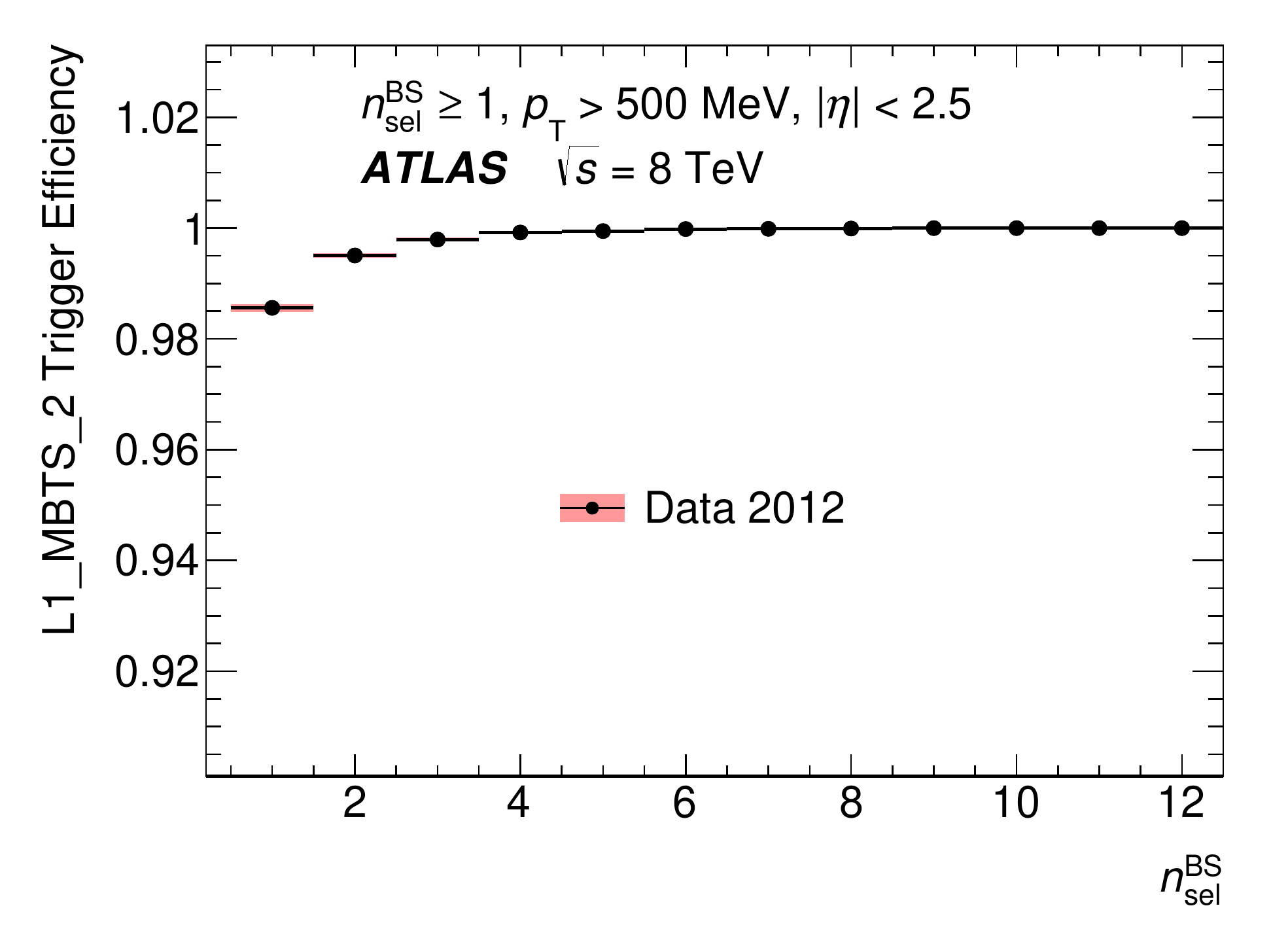}}
    \caption{MBTS trigger efficiencies for $8 \TeV$ data in the two most inclusive measured phase spaces: 
(a) L1\_MBTS\_2 trigger efficiency as a function of the number of selected tracks $\nselbs$ in the phase space with $\nselbs \geq 2$ and $\pT > 100 \MeV$. 
(b) L1\_MBTS\_1 trigger efficiency as a function of the number of selected tracks $\nselbs$ in the phase space with $\nselbs \geq 1$ and $\pT > 500 \MeV$. 
(c) L1\_MBTS\_2 trigger efficiency as a function of the number of selected tracks $\nselbs$ in the phase space with $\nselbs \geq 1$ and $\pT > 500 \MeV$. 
L1\_MBTS\_1 and L1\_MBTS\_2 are the requirements that in at least one or two modules of the minimum-bias trigger scintillators a signal above threshold was registered, respectively. 
The shaded areas represent the sum of systematic and statistical errors.
    \label{fig:auxtrig}}
\end{figure}

The efficiency of the trigger requirement in which the signal in at least one of the MBTS modules was above threshold is presented in \Fig{\ref{fig:trigeff_500_1}} for the phase space with $\nch \geq 1$ and $\pT > 500 \MeV$. 
The efficiency of the trigger requirement in which the signal in at least two of the MBTS modules was above threshold is presented in \Fig{\ref{fig:trigeff_100_2}} for the phase space with $\nch \geq 2$ and $\pT > 100 \MeV$, and in \Fig{\ref{fig:trigeff_500_2}} for the phase space with $\nch \geq 1$ and $\pT > 500 \MeV$. 
All results are shown as a function of the number of selected tracks per event, \nselbs. 
In the phase space with the higher transverse momentum threshold, the trigger efficiency is higher and rises more quickly to \SI{100}{\percent} for both trigger requirements.

\clearpage

\section{Additional vertex and track reconstruction efficiency plots}
\label{sec:morevtxtrkeffplots}

\begin{figure}[htbp]
\centering
    \subfloat[\label{fig:vrtxeff_500}]{\includegraphics[width=0.48\textwidth]{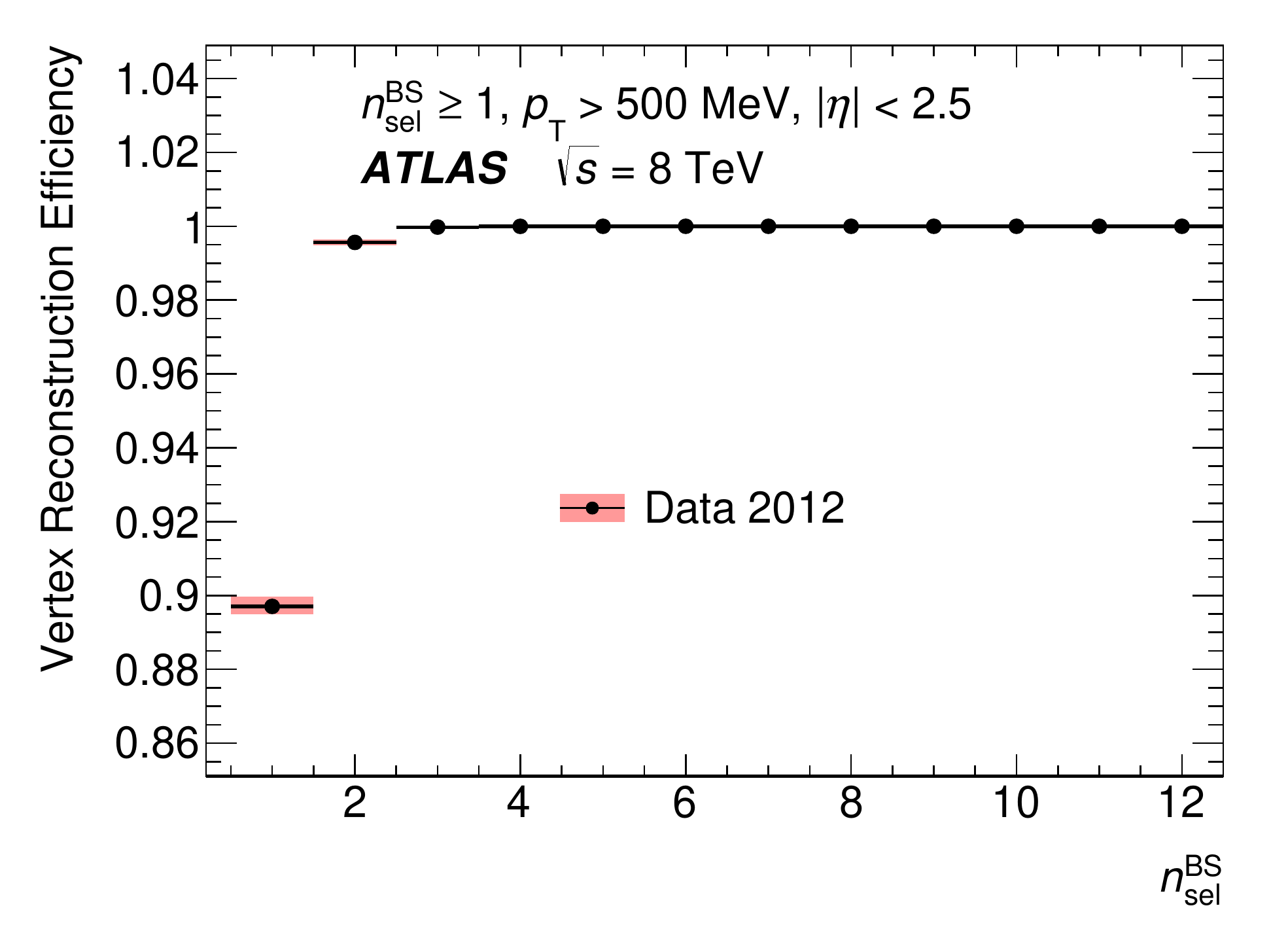}}
    \subfloat[\label{fig:trkeff_500_eta}]{\includegraphics[width=0.48\textwidth]{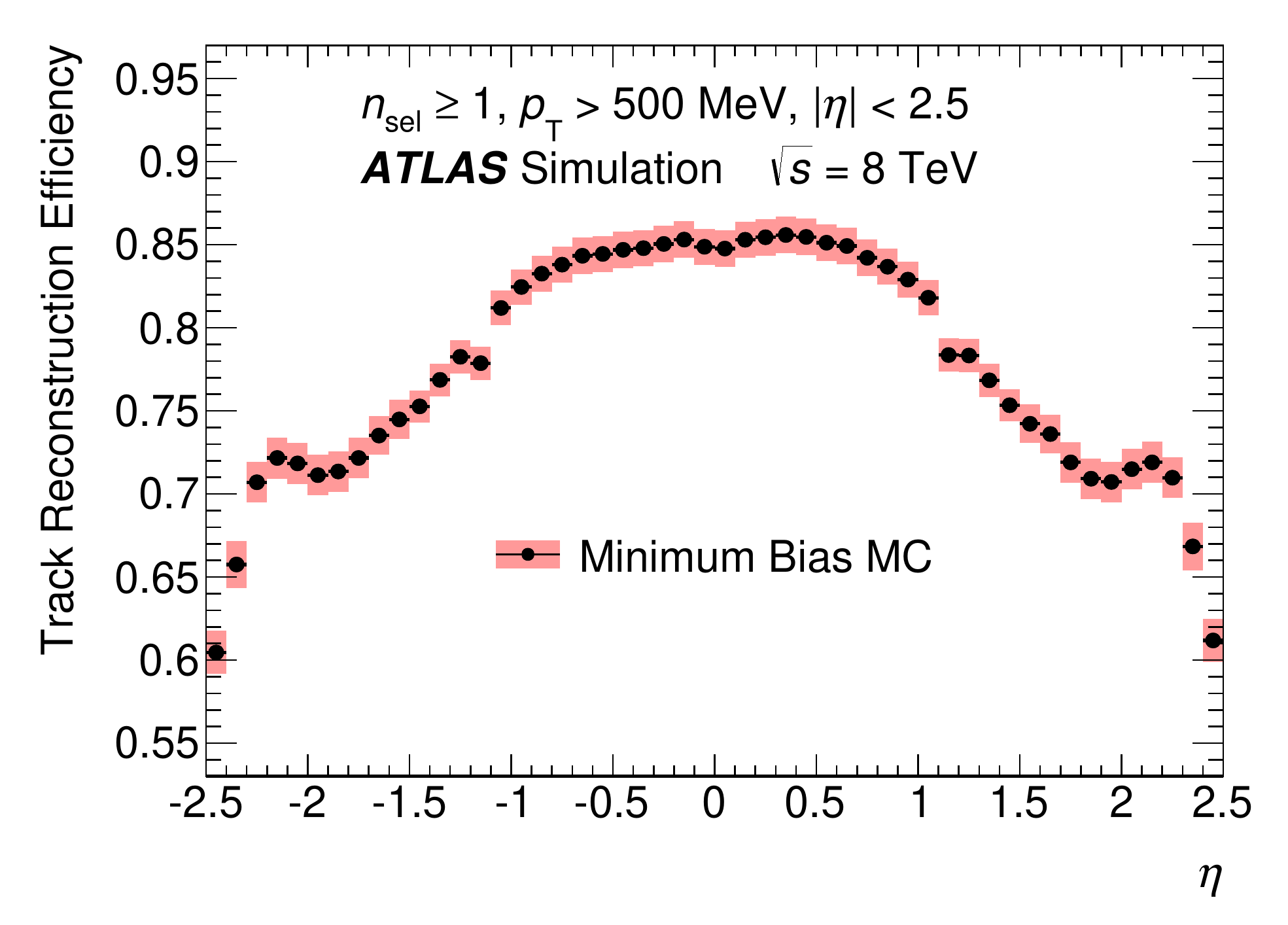}}
    \caption{Selection efficiencies for $8 \TeV$ data in the $\pT > 500 \MeV$ phase space:
(a) The vertex reconstruction efficiency as a function of the number of selected tracks, \nselbs. 
(b) The track reconstruction efficiency as a function of the pseudorapidity, $\eta$. 
The shaded areas represent the sum of systematic and statistical errors.
    \label{fig:auxvrtx}}
\end{figure}

The vertex reconstruction efficiency for events selected in the phase space with the higher transverse momentum threshold, $\pT > 500 \MeV$, is presented in \Fig{\ref{fig:vrtxeff_500}} as a function of the number of selected tracks per event, $\nselbs$. 
The systematic uncertainties are found to be small in comparison with those in the most inclusive phase space. 

The track reconstruction efficiency for tracks from events selected by the $\nsel \geq 1$ and $\pT > 500 \MeV$ requirement, which was determined from MC simulation using the \PYTHIAatwo tune, is presented in \Fig{\ref{fig:trkeff_500_eta}} as a function of the pseudorapidity.

\printbibliography


\newpage 
\begin{flushleft}
{\Large The ATLAS Collaboration}

\bigskip

G.~Aad$^\textrm{\scriptsize 86}$,
B.~Abbott$^\textrm{\scriptsize 113}$,
J.~Abdallah$^\textrm{\scriptsize 151}$,
O.~Abdinov$^\textrm{\scriptsize 11}$,
B.~Abeloos$^\textrm{\scriptsize 117}$,
R.~Aben$^\textrm{\scriptsize 107}$,
M.~Abolins$^\textrm{\scriptsize 91}$,
O.S.~AbouZeid$^\textrm{\scriptsize 137}$,
N.L.~Abraham$^\textrm{\scriptsize 149}$,
H.~Abramowicz$^\textrm{\scriptsize 153}$
% ...
\bigskip
\\
% ...
$^{11}$ Institute of Physics, Azerbaijan Academy of Sciences, Baku, Azerbaijan\\
% ...
$^{86}$ CPPM, Aix-Marseille Universit{\'e} and CNRS/IN2P3, Marseille, France\\
% ...
$^{91}$ Department of Physics and Astronomy, Michigan State University, East Lansing MI, United States of America\\
% ...
$^{107}$ Nikhef National Institute for Subatomic Physics and University of Amsterdam, Amsterdam, Netherlands\\
% ...
$^{113}$ Homer L. Dodge Department of Physics and Astronomy, University of Oklahoma, Norman OK, United States of America\\
% ...
$^{117}$ LAL, Univ. Paris-Sud, CNRS/IN2P3, Universit{\'e} Paris-Saclay, Orsay, France\\
% ...
$^{137}$ Santa Cruz Institute for Particle Physics, University of California Santa Cruz, Santa Cruz CA, United States of America\\
% ...
$^{149}$ Department of Physics and Astronomy, University of Sussex, Brighton, United Kingdom\\
% ...
$^{151}$ Institute of Physics, Academia Sinica, Taipei, Taiwan\\
% ...
$^{153}$ Raymond and Beverly Sackler School of Physics and Astronomy, Tel Aviv University, Tel Aviv, Israel\\
% ...
\end{flushleft}




%

\end{document}